# Microblog Analysis as a Programme of Work [1]

PETER TOLMIE, Universität Siegen
ROB PROCTER, University of Warwick
MARK ROUNCEFIELD, Lancaster University
MARIA LIAKATA, University of Warwick
ARKAITZ ZUBIAGA, University of Warwick

Inspired by a European project, PHEME, that requires the close analysis of Twitter-based conversations in order to look at the spread of rumors via social media, this paper has two objectives. The first of these is to take the analysis of microblogs back to first principles and lay out what microblog analysis should look like as a foundational programme of work. The other is to describe how this is of fundamental relevance to Human-Computer Interaction's interest in grasping the constitution of people's interactions with technology within the social order. Our critical finding is that, despite some surface similarities, Twitter-based conversations are a wholly distinct social phenomenon requiring an independent analysis that treats them as unique phenomena in their own right, rather than as another species of conversation that can be handled within the framework of existing Conversation Analysis. This motivates the argument that Microblog Analysis be established as a foundationally independent programme, examining the organizational characteristics of microblogging from the ground up. We articulate how aspects of this approach have already begun to shape our design activities within the PHEME project.

• Human-centred computing→Human computer interaction (HCI)   • Collaborative and social computing→Social media.

Additional Key Words and Phrases: Conversation analysis, ethnomethodology, turn-taking systems, microblogs, Twitter, rumor, annotation.

## 1. INTRODUCTION

This paper has two primary objectives: 1) to constitute Microblog Analysis as a foundational programme of work; 2) to further articulate how this is of foundational relevance to Human-Computer Interaction and related matters of understanding how people use social media; in particular, how people are seen to attend to the challenges of making the use of social media orderly.

This work has been motivated by the PHEME[2] project and its central interest in the detection of rumors in social media and their subsequent assessment and handling according to the veracity or otherwise of the information they are disseminating [Bontcheva et al., 2015; Zubiaga et al., 2015a; Zubiaga et al., 2016a]. A core part of this endeavor has been the identification of rumorous tweets on Twitter and the annotation of those tweets in terms of features that might ultimately lend themselves to system recognition and machine learning. It seemed reasonable to us

[1]     The research reported in this paper is supported by EC FP7-ICT Collaborative Project PHEME (No. 611233). www.pheme.eu

        Authors' addresses: P. Tolmie, Universität Siegen, Wirtschaftsinformatik und Neue Medien, Kohlbettstr. 15, 57068, Siegen, Germany. R. Procter, M. Liakata, A. Zubiaga, Department of Computer Science, University of Warwick, Coventry, CV4 7AL, United Kingdom. M. Rouncefield, School of Computing and Communications, Lancaster University, Lancaster.

[2]        http://www.pheme.eu/

that, if you are going to try and handle tweets in this way, you had better first of all understand what kinds of things you are dealing with as human socially-constituted phenomena. To do this, we initially brought to bear the analytic apparatus of Conversation Analysis as first laid out by Harvey Sacks and his colleagues in the 1960s and 70s. We did this because it eschews pre-theoretical judgments regarding what kinds of phenomena one might be looking at and instead seeks to uncover empirically how talk-based phenomena are the methodical production of the parties to that production. In this way, we figured we might be able to bring out the methodical features of tweets that provide for their character as rumors (or anything else) in social interaction and that these methodical features would give a handle on what one might want to label when engaged in annotation [Zubiaga et al., 2015a]. However, it quickly became apparent to us that to just apply the apparatus of conversation analysis and seek to identify phenomena already identified within its canon was not wholly satisfactory. Tweeting, for all of its conversational characteristics that people are happy to point to [Honeycutt & Herring, 2009], is not conversation.[3] So, this paper unpacks some of the important distinctions between conversation and 'microblogging' (to take the term that has been applied to a variety of very similar forms of interaction using social media), in order to outline what a suitable approach to analysing microblogging might look like. Building on this, we propose a programme of analysis that is better suited to microblogging. To conclude we demonstrate some of the ways it might then be used by looking at how we ourselves have begun to use it to handle Twitter threads that potentially incorporate rumors. The importance of this particular application we briefly outline below.

## 2. RUMORS IN SOCIAL MEDIA

Social media such as Twitter provide a constant flow of information that is used by many as a news source, especially in cases of emergency situations or at the start of an event, when traditional media have not yet been able to deploy reporters on the ground. The value of social media in coping with the aftermath of natural disasters is well documented [Bruns et al., 2013; Tonkin et al., 2012]. Equally, events such as the 2011 Arab Spring have been heralded as evidence of how social media can strengthen the capacity of citizens to challenge and overcome social and political repression [e.g., Khondker, 2011], though even materials published prior to those events would suggest we treat such claims with caution [e.g. Weaver, 2010]. Journalists and analysts of various backgrounds now monitor social media to identify new stories or gain insights on events unraveling or other areas of interest. Moreover, social media provide a mechanism for people to broadcast their own viewpoint and thoughts, constituting a powerful means for individuals to exercise influence and even mobilize crowds [Procter et al., 2013a; Khondker, 2011].

However, the advent of streaming information broadcast by a multitude of sources comes with one large caveat, that of being able to establish the veracity of a statement, distinguishing between a corroborated fact and a piece of unverifiable information. Indeed, research suggests that social media provides an extremely fertile ground for rumors and misinformation [Mendoza et al., 2010], especially during crises, when unverified statements may sometimes be picked up and given credibility by mainstream media reporting or government agencies such as the emergency services [Procter et al., 2013b]. During an earthquake in Chile, for example, rumors spread through Twitter that a volcano had become active and there was a tsunami warning in Valparaiso [Mendoza et al., 2010]. Twitter has also been used to spread false rumors during election campaigns [Ratkiewicz et al., 2011]. The challenges communities face from rumors in social media is well-illustrated by the 2011 riots in England, which began as an isolated incident in Tottenham, London, on the 6th of August but subsequently quickly spread across London and to other cities in England [Lewis et al., 2011; Procter et al., 2013a].

---

[3]        We note that Twitter's user interface and tweet metadata have evolved in such a way to suggest that its conversational features have become steadily more significant.

It is this problem of verifying information posted in social media that has motivated the PHEME project [Derczynski et al., 2015]. In this paper we focus on closely examining Twitter as a socially constituted phenomenon. This has been a necessary step towards developing an annotation scheme [Zubiaga et al., 2015a; 2015b] for tweets that the project will be using to train natural language processing and machine learning techniques that will assist in the rumor verification process. Whilst it may be tempting to think that one can simply find a way of looking at isolated tweets and see within them already the necessary constituents that might make them a rumor, individual tweets are *made* into rumors by people and the ways in which they are responded to, articulated and spread. These are social processes through and through and can only be understood by understanding the social order that underpins them. To unravel the social order in play one needs the right kinds of tools. Our starting point was Conversation Analysis because it sets aside any theoretical preconceptions regarding the phenomena in play and examines the ways in which social phenomena are constituted in situ through the sequential production of inter-related utterances, which would seem to capture what Twitter-based conversations, rumorous or otherwise, look like[4]. However, there is a risk involved in taking Conversation Analysis as a frame. Doing so rests upon an assumption that tweeting works the same way as conversation. This was not an assumption we could comfortably make. Rather, it seemed important to understand tweeting on its own terms. We therefore looked instead to the foundational insights that first inspired Conversation Analysis as an enterprise to see if this could provide us with insights as to how to proceed when handling how people use Twitter. It is the outcome of that exploration that we present in this paper.

### 3. TWITTER, SOCIAL COMPUTING AND THE CONVERSATION ANALYTIC FRAME

Twitter is a microblogging site that was set up in 2006, which allows users to post messages ('tweets') of up to 140 characters in length. Unlike social media platforms such as Facebook, Twitter's friendship model is directed and non-reciprocal. Users can follow whomever they like, but those they follow do not have to follow them back. When one user follows another, the latter's tweets will be visible in the former's 'timeline'. It is not necessary, however, to follow another user to access tweets: Twitter is an open platform, so by default tweets are public and can be discovered through Twitter search tools. The one exception to this is the *direct message* (DM), which is private, and can be seen only by the follower to whom it is sent. Users can also reference another user through the *mention* convention, where a user name, prefixed with '@', is included anywhere in a tweet. A user, thus referenced, will see the tweet in their timeline. A user can also opt to make their account private, in which case the user can approve who would be able to read their tweets.

---

[4]            Paulus et al (2016) systematically review 89 other CA-based papers that rest upon the same kind of premise. A key part of what we seek to do here is to move the analytic focus beyond CA.

A number of other important conventions have emerged as Twitter use has evolved. One is the retweet option, usually referred to as *RT*, whereby users can forward tweets from other users to their own followers. This works by either clicking on the retweet button available on the standard Twitter user client, or by copying the original tweet and putting 'RT @username' in front of it, the latter giving the option to accompany the original tweet with their own comment. In this way, tweets can propagate through users' follower networks. Another convention is the *hashtag*, which is distinguished by prefixing a string of text with the hash sign, '#'. Hashtags provide a way for users to assign a label to a tweet, thereby enabling the co-creation of a fluid and dynamic thread within the timeline that facilitates information discovery: anyone searching for or using the same hashtag can see what everyone else is saying about this topic. Yet another convention is the *reply* option, which, as the name suggests is a mechanism for responding to a specific tweet. As such, it is a specialisation of the mention convention in that the username of the poster of the tweet being replied to is prepended to the new tweet. Unlike a mention, however, only users (other than the sender and the recipient) who follow both the sender and the recipient will see it in their timeline. Finally, *favoriting* a tweet is a way of letting the tweet's poster know that you liked the tweet. The usage patterns of each of these conventions differ and in the case of hashtags there is recent evidence that their appeal is diminishing [Rahimi, 2015].[5]

If one looks at studies of Twitter both within and beyond social computing what one finds in abundance are studies that look in one way or another at the *content* of tweets and how that content might be seen to relate to a variety of other matters. Thus, and for instance, we find treatments of the content of Twitter-based conversations in the course of political campaigns, or in other political contexts. As an example, Burgess & Bruns [2012] looked at the content of tweets associated with the #ausvotes hashtag during the Australian elections in 2010 to examine the character of political discussion on Twitter. In particular, they provide examples of tweets to demonstrate that the most popular tweets were not engaging at all with any of the mainstream political reporting (see Stieglitz & Dang-Xuan [2012] and Jungherr & Jurgens [2014] for other examples of this kind of tweet-based analysis). Equally common are examinations of Twitter posts in the aftermath of natural disasters, or as a feature of other kinds of emergencies or security scares. Chatfield et al. [2014], for example, examine the way in which government bodies made use of Twitter in the aftermath of Hurricane Sandy. Additionally, they provide examples of tweets to evidence how citizens were using Twitter to circulate specific points of advice. This is built into an argument that citizens have a critical role to play in supporting the public services under such circumstances. (See also Mandel et al., [2012] and Purohit et al. [2013] for other treatments of the same kind of material). Also popular are examinations of the use of Twitter during major news events. For instance, Hu et al. [2012] examine large number of tweets before, immediately after and some time after the news of Osama Bin Laden's death to try and pull out markers of 'certainty' from the content. They note, in particular, the impact politicians and journalists tweeting in their own right were able to have on this score and this is then used to argue for the important role Twitter may play in disseminating breaking news more in relation to more traditional mass media channels. (Other content-based treatments of the handling of news events can be found in: Bruns & Burgess [2012]; Gupta & Kumaraguru [2012]; Procter et al., [2013b]; and Zubiaga et al., [2016a]).

---

[5] It should be noted that the original version of this paper was written in early 2015 and that a number of the conventions discussed here have changed since then. This does not undermine the key points we make in this paper about how Twitter is oriented to and operates as a turn-taking system. What it does emphasise is the extent to which social media platforms such as Twitter are in a constant state of evolution as designers and service providers continue to look for new kinds of functionality or seek to respond to issues raised by their clients (who encompass more than just individual users, of course). This makes social media something of a moving target for research, and strengthens the case for undertaking foundational research that captures how social media systems work as organisational phenomena at a methodological level, rather than just in terms of their content. We would like to thank the reviewers of this paper for first bringing this to our attention.

Other bodies of tweet content examined include those used in specific settings such as conferences, learning environments, or the workplace. Gonzales [2014], for instance, looked at when and what tweeters typically tweet about whilst attending conferences, using content from a range of tweets to make the case that Twitter plays an important role in supporting discussion at conferences. The presentation of tweet content to make a case for how Twitter may support specific kinds of interactions amongst certain communities can also be found in Borau et al. [2009], and Bougie et al. [2011], Hambrick et al. [2010], and Morris [2014].

Yet another sort of content analysis has been devoted to the use of Twitter in different kinds of interpersonal communications and relationships. McPherson et al. [2012] looked at how people were tweeting one another whilst they were actually engaged in watching the television show *Glee* to see what part this might be playing in the shaping of their viewing as a social experience. Specific tweets were gathered and categorized according to whether they were comments about characters, general comments about the show, comments about the episode they were currently watching, and so on. Associated interviews and analysis were then used to bring into question whether these kinds of interchanges amount to actual conversations. Examples of this kind of focus on Twitter abound. Bak et al. [2012] looked at how certain kinds of content in Twitter conversations might be indicative of self-disclosure and the strength of relationship between the participants; Murnane & Counts [2014] examine the Twitter posts from smokers and the part they play in them giving up smoking; Kendall et al. [2011] examine how people use Twitter to share information with one another about their health and fitness activities; Magee et al. [2013] use tweets amongst players of a role-playing game to explore the part played by these in enhancing their gaming experience; Zhang et al. [2011] look at how consumers use 'word-of-mouth' on Twitter to share information about businesses and their products and services; and Sleeper et al. [2013] examine a range of tweets that tweeters have subsequently regretted posting and the actions they have then taken to engage in some kind of repair.

Another body of work can be seen to be stepping beyond the cherry-picking of 'interesting' content in order to take the actual constitution of tweeting as *conversation* as its topic of interest. Cogan et al. [2012] use a graphing approach to try and extract whole Twitter conversations given a conversational root. Larodec et al [2014], by contrast, use content analysis to try and identify user interactions on Twitter as a way of getting at conversations. On a different tack, Schantl et al. [2013] attempt to develop a model that can be used to predict who the repliers might be in Twitter-based conversations. A particularly significant trend is the quest to identify trending or persistent topics. This may involve: looking at a large number of trending topics to try and understand how they first arose and the dynamics through which they became popular (Ferrara et al. [2013]); examining how topics within tweets might actually be identified (Inches & Crestani [2011]); or looking at how to identify specific topics that are trending at specific moments in time (Shamma et al. [2011]).

Conversation focused work also includes examination of how to identify and model specific kinds of conversational acts. In this vein, Huang et al. [2010] look at the specific practice of adding hashtags to tweets and the kinds of things it is designed to accomplish; whilst Naaman et al. [2010] consider how to use the content of tweets to identify different kinds of user behaviour; and Ritter et al. [2010] look at how interactions between users on Twitter might be modeled as 'dialogue acts'. Research has also been devoted to matters of conversational address and coherence, and how Twitter conversations are topically organized. Honeycutt & Herring [2009], for instance, have looked at the use of the @ sign as a marker of address; Lai & Rand [2013] have examined how topics of conversation unfold dynamically; and Sommer et al. [2012] have focused upon both topic relations and sentiment in Twitter conversations. Broader approaches have also included reflection upon how to analyse the structure of tweets in terms of conversational frameworks. de Moor [2010], for instance, developed a socio-technical context framework to examine how the technical aspects of Twitter were related to the production of conversations. At the same time Kumar et al. [2010] were exploring the structure of Twitter conversations to establish whether there was scope for developing a mathematical model of how they were organized. Relatedly, Zappavigna [2012] examined how Twitter interactions might be formulated as specific forms of electronic discourse.

Other authors concern themselves more with how to analyse Twitter- and microblog-based phenomena in their own right, for instance, by looking at the use of specific features such as hashtags (Laniado & Mika [2010]), or retweets (Boyd et al [2010]). Other considerations look at Twitter's linguistic character such as the use of curses (Wang et al. [2014]), or its multilingual character and how it might itself constitute a language ecology (Eleta [2012]).

Quite a different set of analyses have looked at the social character of Twitter. Rossi & Magnani [2012] look at how the public nature of Twitter has an impact upon the ways in which people use it. Schoenebeck, [2014], on the other hand, has looked at how people take breaks from using Twitter and their reasons for doing so, and Stibe et al. [2011] have examined how Twitter can be used as a tool for persuasion and the influence some users may have upon the attitudes of others. Yet another set of studies focus upon Twitter as a vehicle for information-exchange and matters such as the identification of people's information requirements by examining how they use Twitter (Zhao & Mei [2013]) and the extent to which information needs may not currently be well-supported (Ramage et al. [2010]). This, of course, also relates to our own interest in rumor, with Qazvinian et al. [2011], Starbird et al. [2014], Maddock et al [2015], and Zhao et al. [2015] all using a range of techniques to try and detect the presence of rumors in tweets and to then relate this to rumor diffusion.

Other studies again have looked at Twitter use in terms of motivation or interest. By way of example, Alonso et al. [2013] used crowdsourcing techniques to try and identify what kinds of content people find interesting on Twitter; and both Naveed et al. [2011] and Azman et al. [2012] looked at what kinds of motivation people might have for retweeting.

In view of the scale of the literature the above is necessarily a very cursory overview of some of the principle ways in which scholars have tried to approach the investigation of Twitter-based phenomena. However, what all of the above approaches engage in at some level is an assumption that we all already know what Twitter 'is' as a social phenomenon (and, in a sense, as ordinary users, we already do): it's tweets about threats and troubles; about unfolding events; about who has said what or who is doing what; about things you've accomplished; about things that have piqued your interest; and so on. This is not intended to set aside in any way the significant academic endeavour and intellectual rigor involved in many of the studies we have cited. It is also important to acknowledge the extent to which Twitter has changed in both format and use over the 10 years since its inception, which has itself had an impact upon the kinds of studies and publications addressed to that use. However, we feel that it is a matter of foundational research concern that we should not *a priori* assume that we all just know how to handle Twitter use as a phenomenon. All too often the understanding evinced in current studies is indexed upon a commonsense understanding of what Twitter is about, and what kind of a thing we might want to describe it as, without ever digging into the grounds of that commonsense understanding itself. Thus, just as with the problem Garfinkel [1967] was seeking to address in his early work regarding an overwhelming propensity in social sciences to speak of society but to leave the actual accomplishment of society untouched, we are confronted here with a similar tendency to set the actual accomplishment of tweeting as a social phenomenon aside and to instead simply work with its products, i.e. how the content is used. That is, as Garfinkel might have suggested, actual Twitter use is not investigated or researched as a 'topic' but is instead treated as if we already understand the phenomenon and is used as a mere 'resource' in the investigation of other matters (see Garfinkel & Wieder [1992]). Thus there is an ongoing absence in the literature regarding the nature of the technology[6] that is being brought to bear and its impact on social interaction and what that interaction therefore looks like.

Taking seriously the early work of Sacks, Schegloff and Jefferson [1974] regarding just what the organizational properties of conversation might look like as an effective system for getting the job of co-situated talk done, we argue here that there is a similar need with Twitter and other like phenomena (characterized here as microblogging) to go back to basics and look at their organizational characteristics and that this is the only effective way of being able to handle them as social phenomena.

It's some time now since ICT systems design first took its 'turn to the social' (see Crabtree et al. [2012] and Button et al. [2015]), famously instantiated in the works of Suchman [1987] and Grudin [1990]. As pointed out in Button et al. [2015] a part of this turn was what might be seen as a rather problematic flirtation with social science *theory* – postmodernism, post-feminism, queer theory as well as the cultural, linguistic and textual 'turns' [Bardzell & Bardzell, 2011; Light, 2011; Rode, 2011] and so on – where it was assumed that this could simply be imported wholesale from social science into systems design. It is not our position here to critique the progress or results of this frequently less than happy, or productive, interdisciplinary endeavour, other than to point to Anderson and Sharrock's comment that: *"the alignment of sociological theory and design specification remains intractable ... designers and especially members of the HCI research community have continued to advocate incorporation of forms of social and sociological theory into design but with very little substantive success"* [Anderson & Sharrock, 2013]. However, another and perhaps rather more fruitful part of 'the turn to the social' was an equally willing embrace of social science *methods* to reveal, document and evaluate aspects of user experience etc. Conversation Analysis was, of course, one of these methods and its use has, in fact, played a quite significant role in social computing over the years (see, for instance, Heath & Luff's [1991] analysis of interaction in control rooms for the London Underground and Ruhleder's [1999] analysis of communication breakdowns in video-mediated communication across remote sites). We would also point out that another strand of 'social' research in HCI that has been of some moment is an attention to the use of technology for the production of text in interaction, such as Grintner and Eldridge's work on the use of SMS by teenagers [2001] and Curtis's work on social interaction in MUDs [1992]. This paper can therefore also been seen as a continuation and development of these kinds of lines of enquiry, considering the intimate connections between the technology and talk or text in social interaction, but with a more specific and detailed emphasis upon a proper consideration of exactly what an appropriate methodology for such undertakings might need to look like.

---

[6]        We use technology here in its grandest sense, incorporating not just the computing technology required for its production but also the technical apparatus whereby such interaction might get done, just as the turn-taking system for conversation outlined by Sacks et al. [1974] is a technical apparatus for getting talk done, even if, on occasion, it might involve the use of specific technologies such as the telephone.

This being the case, we set aside here the assumption that we know already how to analyze Twitter feeds, even if we regularly process such content both as users and researchers. We similarly set aside the assumption that tweeting is just conversation, even if conversation is a label that is often apparently convenient to use. Instead we set about here trying to establish from the ground up what an appropriate framework for analyzing Twitter feeds might look like, and how that is in fact quite distinct from the conversation analytic enterprise that we first thought we might use. To this end, from this point onwards instead of using the term 'Twitter-based 'conversations'' we shall use the more general term Twitter-based 'interactions', holding in abeyance at this stage any assumption that we know what these interactions amount to.

### 4. METHODICAL PRACTICES AS SOLUTIONS TO ORGANISATIONAL PROBLEMS

Harvey Sacks, in some profoundly significant remarks regarding the way in which any orderly organizational apparatus geared towards the accomplishment of a coherent social order would have to operate, noted that such an apparatus needs to be available to just any member of society such that they could make use of it without much fuss or bother or the need to engage in extensive formal training or the accumulation of multiple examples of its use. It is in the spirit of these remarks that can be seen to inhabit the seminal work he undertook along with Emmanuel Schegloff and Gail Jefferson in specifying some of the fundamental organizational characteristics of the turn-taking system in conversation [Sacks et al., 1974].

What can be seen in this work is a recognition that the bringing about of a particular aspect of social order takes, first of all, seeing that order as a 'problem' that requires the application of a method to be addressed. The notion of order being a problem is not to set aside its mundane and wholly unremarkable character, to be found for the larger part wherever one looks. Indeed, the very sense of it being unremarkable is itself an accomplishment. Instead, the approach is one of seeing that order does not just arise as if by magic wherever people chose to go and whatever people choose to do. Rather, it is accomplished by people in regular, methodical ways such that they don't have to keep learning new methods to make their way around the world. So, the simplest systematics takes turn-taking in conversation to be an elegant and simple solution to a wholly mundane problem that people are confronted with whenever they engage in social interaction. If they all talk at once none of them will be understood. The simplest systematics therefore unfolds as an articulation of just what the accomplishment of turn-taking needs to look like to be a workable and coherent system. But what Sacks et al. also accomplish in doing this is the demonstration of a *programmatic* approach to the description of a social order that is, even up until now, largely undescribed.

In that our current work seeks to establish microblog analysis as a programme of work, it seeks to proceed along the same lines as the simplest systematics by taking, in the first instance, the phenomena associated with microblogging to be methodically constituted solutions to arising organizational problems in the work of communicating in that way. In that case it will be seeking a) to uncover just what those organizational problems might be and b) to describe in methodological terms how microblogging phenomena represent ways in which those problems are being recurrently addressed.

As a note of clarification, the term microblogging has been adopted here in order to capture a set of rather similar communicational phenomena to be found across a range of social networking sites and in certain kinds of forums. The original term used to cover these kinds of short text-based phenomena was 'tumblelogs' [Kottke, 2005], but by 2006 'microblog' had become the preferred term. It was used to cover a variety of services such as Twitter, Tumblr, FriendFeed, Cif2.net, Plurk, Jaiku, identi.ca, PingGadget and Pownce [Barnes & Bohringer, 2011; Honeycutt & Herring, 2009; Huang et al., 2010; Java et al., 2007; Kaplan & Haenlein, 2011; Lai & Rand, 2013; Naaman et al., 2010; Oulasvirta et al., 2010; Zappavigna, 2011; 2012]. More recently the term has come to also cover the status update features of social networking sites such as Facebook, MySpace, LinkedIn, Diaspora*, JudgIt, Yahoo Pulse, Google Buzz, Google+ and XING [Archambault & Grudin, 2012; Chen & She, 2012; Gilbert et al., 2013; Larodec et al., 2014]. The specific worked through case here is that of Twitter.

More than this, our focus here is upon interactions using Twitter that are now recurrently grouped by Twitter itself into 'conversations'. It is, of course, the case that Twitter feeds, can take many forms, and many postings to Twitter stand as isolated fragments of text or images, that are not produced as a direct response to other postings, and that are not productive of other responses in their own right. This alone puts much of the content of Twitter outside of anything Conversation Analysis might seek to tackle. Our interest in examining the production of Twitter interactions, where there is an interlinking of related texts, derives from our interest in the propagation of rumors on Twitter. The spreading of rumors on Twitter necessarily entails posts being productive of further postings, otherwise the phenomena would not be describable as rumors in the first place. Thus it is the case that in this paper we are only addressing part of a much larger enterprise that would seek to understand what tweets and tweeting might amount to as organized social phenomena across the board rather than just in the context of Twitter interactions. Nor is there any assumption here that what can be said for Twitter can be said for all in all regards and significant further work would need to be undertaken in each separate microblogging domain to fully capture their organizational characteristics. By focusing on Twitter-based interactions, however, we found that, just as is the case with the materials first gathered for exploring the organizational character of conversation (see Sacks [1984]) we were confronted with a turn-taking system and proceeded to examine it accordingly.

Thus our focus upon 'microblogging' here is designed to capture a specific kind of text-based exchange where contributions to the exchange are relatively short and constrained. They are articulated within textual confines even if they contain other kinds of media, produced with the prospect (if not the expectation) of others being able to respond, and with related contributions to the exchange being open to being produced asynchronously. They are also produced, at least in the first instance, in a single largely undifferentiated stream that is temporally organized with the latest contribution being placed at the top of the list. This description provides for the possibility of a variety of social networking sites being analyzed in a similar fashion.

## 5. MICROBLOG INTERACTIONS AS A FORM OF TURN-TAKING

As articulated above our prime interest here is in understanding how microblog exchanges on Twitter work as human organizational phenomena. As we also point out above, the exchange format of Twitter interactions makes them, operationally at least, a species of turn-taking. What we mean by this is that the exchanges in question are made up of a series of inter-related parts, where one part is implicative of another, and where the different parts are produced by at least two different parties. Thus the job of understanding how these exchanges work entails, amongst other things[7], looking at how the parts are organized in relation to one another as a series of turns. Now, fortunately for us, we are not the first to examine how identifiably distinct utterances might be organized as a system of turn-taking. A central work in the conversation analysis canon, *A Simplest Systematics for the Organization of Turn-Taking for Conversation*, by Harvey Sacks, Emmanuel Schegloff and Gail Jefferson [1974], undertook exactly this job of work when examining the organizational properties of face-to-face conversation. We will therefore be making use of a number of their key insights here in order to articulate precisely the ways in which Twitter-based exchanges are phenomenologically *distinct* from ordinary face-to-face conversation.

At the very heart of Sacks et al.'s *Simplest Systematics* is the observation that talk is organised such that only one speaker speaks at once. This is seen as a fundamental premise of social order because any other system would frequently render talk completely ineffectual. On the basis of this, and probing just how it could be that this is systematically provided for in interaction, Sacks et al. elaborated what they called the 'turn-taking mechanism'. They saw this system as containing some primary features that together serve to underpin most other kinds of conversational phenomena. Thus, and for instance, there are: speakers (recognizable individuals who produce utterances); speakers who talk first, and other speakers who may also talk as a conversation unfolds; mechanisms whereby a current speaker may select who talks next; and mechanisms whereby speakers may select themselves to be the next person to produce an utterance.

They also made a point of noting that human action and interaction is replete with examples of turn-taking, so:

> "Turn-taking is used for the ordering of moves in games, for allocating political office, for regulating traffic at intersections, for serving customers at business establishments, and for talking in interviews, meetings, debates, ceremonies, conversations etc." [Sacks et al., 1974: 696].

A number of online activities that were not a feature of human interaction at the time may be seen as further candidate members of this list, including microblogging.

---

[7]        We make no claim to comprehensiveness in our analysis here. This is merely a first foray and there are a number of organizational properties present in Twitter exchanges that we are not examining here, e.g. the relational order holding between different parties; the situated order holding for each productive party beyond what is visible in the timeline; the historical and prospective character of tweet production as a reasoned feature of a specific tweeter's assemblage of tweets; and so on.

A key justification of their work that is of equal moment here was the following: *"it is of particular interest to see how operating turn-taking systems are characterizable as adapting to properties of the sorts of activities in which they operate"* [ibid]. Thus *"an investigator interested in some sort of activity that is organized by a turn-taking system will want to determine how the sort of activity investigated is adapted to, or constrained by, the particular form of turn-taking system which operates on it."* [ibid]. It is this spirit of trying to understand the organizing properties of tweet exchange in Twitter as a system in its own right that motivates this current body of work.

Whilst we will be using the *Simplest Systematics* to compare certain aspects of turn-taking in Twitter to turn-taking in conversation it is important to note here that our aim is not simply to apply the *Simplest Systematics* to microblogging in Twitter. Instead our interest is based upon how the *Simplest Systematics* may provide a *model* for how one might proceed to examine turn-taking systems and uncover their organizational features[8]. As a model the *Simplest Systematics* can be seen to have a number of key strengths. One of the most important aspects of all is that the proposed model is able to be simultaneously 'context-free' but also exceptionally 'context-sensitive'. So you can dip into whomsoever, wheresoever and find the same system in play, with the same key operational characteristics. At the same time, the system can be endlessly adapted to meet the particularities of local need without having to step outside of the system itself [Sacks et al., 1974: 700].

### 5.1 Turn-taking in Twitter-based interactions

In order to map out some of the key organizational features of turn-taking in Twitter-based interactions we are going to set out here a range of foundational observations about Twitter that can together be seen to form the building blocks of a thoroughgoing analysis of microblogs that can be expanded upon, developed and refined over time. Where relevant we shall indicate how these observations relate to the observations Sacks et al [1974] made about the organization of turn-taking in conversation in the *Simplest Systematics*.

*5.1.1. Tweeter Change.* One of the core observations that Sacks et al. make [1974: 700] is that the participating parties within the system change over time: it is not just one party doing all the work. In Twitter this is a direct function of who is being followed, the frequency with which they tweet, and the presence of other factors such as promoted tweets. It is conceivable that someone might follow just one other party in which case tweeter change would be rare. However, promoted tweets usually result in some extraneous tweets appearing on anyone's timeline throughout the day. More importantly, in view of the fact that even in 2012 the average number of people being followed for Twitter users was 102 [beevolve.com, 2012] and Twitter has expanded since then, tweeter change is a characteristic of most people's timelines and, for the larger part, two or more tweets concurrently by the same person is infrequent though it certainly occurs.

*5.1.2. Timeline Organization.* As Sacks et al. observe [1974: 700] it is 'overwhelmingly' the case in face-to-face conversation that only 'one party talks at a time'. Now it is the case that in Twitter some of its characteristics are managed as much as anything by its technical configuration and this is one such case. Now, of course, it is in the nature of Twitter that people can be composing tweets at the same time as one another or at widely spaced intervals. However, the actual appearance of a tweet on people's timelines is a function of the time it is registered by the system. Thus, a) whilst it is entirely possible for two parties to be composing tweets in overlap, even in Twitter interactions; b) it is *also* a feature of Twitter that the timeline is organized in independent tweets that do not appear in a simultaneous and overlaid fashion and that do not overlap. Thus the individual actions and their representation are potentially disjoint but this disjuncture is never made visible to recipients.

*5.1.3. Tweet Separation.* Something made much of in the original *Simplest Systematics* is the fact that, whilst turns between contributors in the system are independent, overlaps in turns do, in fact, often occur, though they are typically very brief [Sacks et al., 1974: 701]. However, in the case of tweet exchange, Twitter does not represent overlaps of articulation, even if such overlaps *are* occurring at an individual level. The consequence of this is that within the timeline each turn appears to be tightly independent and consecutive. This has some important implications that we shall be returning to in due course.

*5.1.4. Temporal Disjuncture.* Having pointed out that small overlaps in turns do occur in the case of the conversational system they were examining, Sacks et al. [1974:701] also emphasise that 'transitions (from one turn to a next) with no gap and no overlap are [also] common', and that, together, these three possibilities 'make up the vast majority of transitions'. In the case of Twitter temporal gaps between the appearance of one tweet and another *do* routinely occur. This is often emphatically the case in Twitter interactions where people have to take time to read another tweet, with tweets coming 'all at once', so to speak, because recipients are not able to witness the ongoing composition of the tweet' (something that some messaging services do provide for by the way, e.g. the Beam Messenger for Android phones[9]). Contributors to a Twitter interaction may also come to it some time later: this is another point to which we shall later return. However, all of these temporal disjunctures do not manifest themselves as 'gaps' in the timeline but rather as delays in updates. Once again the extent to which delays in updates occur is tightly bound up with the number of people being followed and the frequency with which they tweet. Broadly speaking, though, temporal disjuncture between tweets can be considered to be a routine feature in Twitter, making gaps in interaction an unremarkable feature of use that is not oriented to by users as problematic or subjected to efforts to repair. This goes hand-in-hand with describing exchange systems like Twitter as being both 'synchronous' *and* 'asynchronous', but it does also immediately render it as something quite distinct from face-to-face conversation[10].

*5.1.5. Turn Order.* A feature of tweet exchange organization that it shares with a number of other turn-taking systems including conversation, is that the actual order of turns 'is not fixed, but varies' [Sacks et al, 1974: 701-702]. In Twitter just where a tweet will fall next in the timeline is not predictable in advance. A notable consequence of this is that one can get the following kind of pattern where a whole set of distinct tweeters all respond individually to an initial opening gambit without any predictable relationship between them (Fig. 1).

---

[9]     www.beammessenger.com

[10]     Of course, as it does not involve co-presence or spoken language, microblogging differs from regular face-to-face conversations in other ways as well. For instance in Twitter there are no available non-linguistic conversational cues, though emoticons and equivalent symbols and abbreviations are sometimes used to do some of the same work, often borrowing heavily upon the established ways of going about this in text messaging. We should also point out that this renders Twitter quite distinct from many other kinds of *online* interaction that people now regularly engage in, such as Skype calls or other sorts of video-conferencing, online classrooms, and so on. Indeed, each of these latter phenomena also merit similarly close investigation in their own terms.

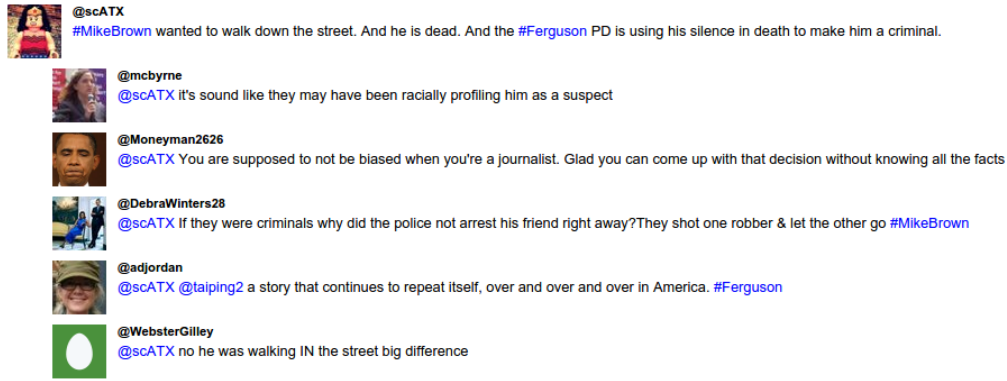

Fig. **1**. A typical timeline.

Another important point we return to later is that this business of turn order not being fixed can result in a potentially indefinite number of people self-selecting to tweet in response to a prior tweet. It would appear that the only systematic constraint in operation here is the size of the cohort of people who follow the person who tweeted initially (with retweeting creating scope for endless extension of this cohort to other users' followers). Looking for a moment outside of Twitter *interactions*, self-selection is clearly entirely routine, and is often occasioned by matters wholly outside of Twitter[11] . This means that it is frequently not clear that parties producing originating tweets are oriented to implicating any overt responses from their followers[12] at all. In the case of the turn-taking system that Sacks et al. [op cit] were looking at, face-to-face conversation, the number of potential self-selecting next speakers is tightly controlled both by the limits that exist on the number of people who can be co-present and in range to hear, and by a range of incumbent rights and obligations that exist as a feature of the relationships that hold between those people who are co-present.

*5.1.6. Implicated Turns and Tweeter Selection (i)*. A fundamental observation to be made here is that, whilst turn order in Twitter is not fixed, there *are* ways in which next turns can be implicated, a feature Twitter interactions do share with other kinds of turn-taking phenomena, and people may be held accountable for the production of turns implicated in these ways. It is therefore the case that there are both self-selection and next tweeter selection techniques in Twitter. Some important differences between this and the system Sacks et al. [1974] were examining, however, are that a) next tweeter selection does not necessarily provide for that next tweeter responding as the next tweeter in the interaction's timeline. Responses can happen after a number of other people have chipped in. Furthermore, b) many self-selected tweets do not implicate continuing interaction anyway. In this case interaction evolves by respondents finding within the tweets the grounds for their own self-selection. So, whilst Twitter may be asynchronous and public facing it is unlike online fora and question answering websites because the goal of a post is not necessarily to start an interaction.

---

[11]         Indeed, the occasioning of first tweets is potentially a whole further job of research that has yet to be undertaken in any thoroughgoing or empirical way. It presents unique research challenges in that it requires situated observation of tweeters in whatever environments they may be happening to tweet from. In this regard, a recent paper by Reeves & Brown [2016] makes a strong case for examining how the production of social media texts and interaction with them needs to be studied as those activities occur, *in situ*, in the ordinary flow of everyday life.
[12]         This is not to say, however, that tweets are not 'recipient-designed' [Sacks, 1992]. On the contrary tweeters remain accountable to their followers for just what they tweet and in what way as we shall examine further on our discussion regarding repair. It should also be noted here that responses from followers in Twitter can take other forms apart from textual replies. One such response is 're-tweeting', another is 'favoriting'. Neither of these have clear conversational parallels; though re-tweeting does have some superficial similarity to reported speech.

*5.1.7. Tweet Length.* An important aspect of Twitter is that, whilst the exact length of a turn may not be pre-specified, its maximum length is very tightly constrained at 140 characters [13]. This puts it in direct contradistinction with the *Simplest Systematics*, where it is observed that turn lengths are not fixed and can vary significantly. Having said this, strategies can be adopted in Twitter that result in something akin to an extension of a turn. One such strategy is the linking of posts (see Fig. 2).

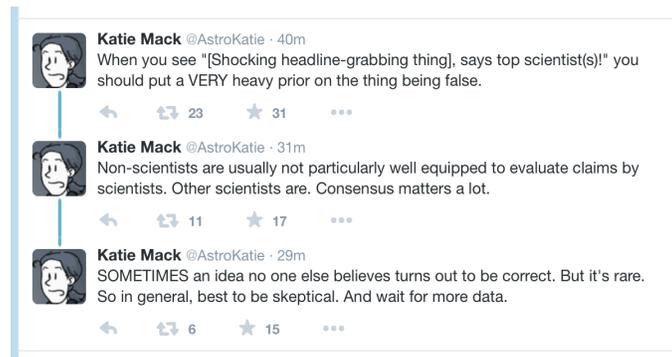

Fig. **2**. Linked posts in a timeline.

Another is the completion of the turn over multiple posts, using the convention of three dots at the end of each post to indicate that there is more to come (see Fig. 3).

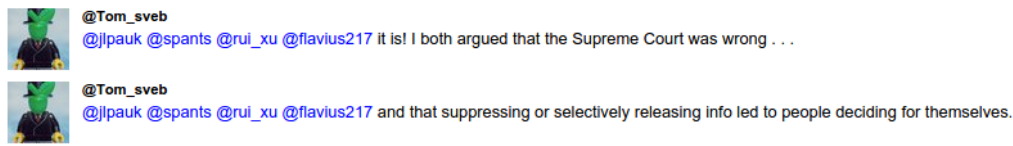

Fig. **3**. Linking with dots.

And yet another way of linking posts is by numbering them, so that it is previously announced how many more posts are coming (Fig. 4).

---

[13]        It should be noted here that, at the time of writing this article, discussions were afoot in Twitter which were actively geared to dispensing with this long-standing feature, which would have potentially led to much greater variety in the length of turns, (Wagner [2016]). Up to 10,000 characters were already being allowed in Direct Messages from 2015. However, following significant public outcry Twitter retrenched from this position, though it has continued to look for other ways in which tweeters can expand their posts, for instance via linking (Spangler [2016]).

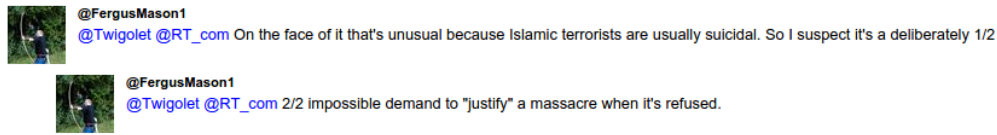

Fig. 4. Numbering posts to link them.

However, it should be noted that in the context of the timeline, as it is encountered by recipients (or, to be more accurate, 'followers'), these strategies still result in separate posts that look to all intents and purposes like separate turns. This needs especially to be borne in mind in view of how tweets like this may appear on a follower's timeline separated out by other posts, even if they are produced consecutively by the tweeter and appear adjacent in the capture of the interaction alone. Fig. 5 shows a way in which the connection can be further emphasized in some cases by including additional elements such 'Cont' before the dots and a reiteration of the last part of the previous post in the subsequent one.

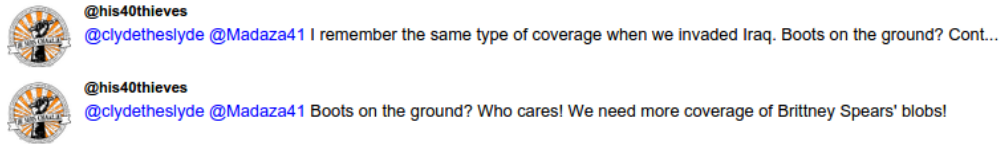

Fig. 5. Additional linking strategies.

It is clear that these strategies are oriented to a fundamental difficulty present in Twitter: maintaining cross-post coherence. In traditional conversation analysis, these strategies would be typically labeled as 'topic reference markers' (Sacks [1992]) (e.g. like 'but as I was saying before', 'with regard to…', 'coming back to…', etc.) in that there is a marker external to the actual content to be provided that makes evident to recipients the presence of a connection. However, in face-to-face conversation these are primarily solutions to problems of temporal disjuncture and topic change. In Twitter they are primarily solutions to the fact that turn-size is pre-specified, or at least absolutely constrained. Thus they are used to extend a turn, not to return to some topic previously mentioned. Having said this, they may, of course, *also* operate as techniques for re-establishing topics which might otherwise have been set aside, even if this is not their most frequent function.

*5.1.8. Duration of Linked Exchange.* An important distinction Sacks et al. [1974: 702] make between naturally-occurring conversation and other kinds of turn-taking talk-based phenomena[14] is that the unfolding talk is not, in principle, constrained to some specific duration. This is not to say that the conversation has no projectible conclusion, no controls over length of turns, no control exercised upon the choice of topic, no constrained rights to speak, etc., but rather to acknowledge that this is something that is locally managed within the course of the talk itself, rather than something that is imposed upon the talk by other external considerations. We can observe here that linked exchanges on Twitter apparently share this characteristic with naturally-occurring conversation, though not with quite the same organizational outcomes in all regards. Thus we have already seen that, in Twitter-based exchanges, the length of a turn and the management of the topic it relates to have some quite distinct organizational characteristics imposed upon them by the nature of the technology and the rigid limit placed upon the number of characters available for use in any one turn.

---

[14]     E.g. pieces of theatre, formal debates, religious rites, and so on.

We shall also see below that the routinely asynchronous character of Twitter has distinct consequences for how such exchanges may unfold. Nonetheless, it remains the case that Twitter-based interactions are not subject to any particular kinds of external constraint upon the amount of time within which they may unfold. More than this, Twitter-based interactions, because of the scope for participants to join and leave the thread over extended periods of time, without any pre-defined preference for temporal contiguity, can unfold over the course of a day or even days as people encounter relevant tweets within their timeline at convenient moments within their own external routine.

Against this, there has to be set another standard technological and organizational feature of Twitter. On its mobile phone-based application Twitter routinely excises tweets from the stream going back more than a couple of hours previously and just glosses their presence with the words 'load more tweets'. Users can click on this and display the missing tweets from the stream, but the fact that they are not displayed as a matter of course has clear implications for what people may most readily engage with. Additionally, because the timeline is relentlessly chronological in its display, at least upon first encounter, going back through the stream to older tweets you may have missed on *any* interface is physically laborious and may involve significant amounts of scrolling down the page to get at them, depending on the number of people being followed, and this too has implications for what may more readily be encountered and acted upon as a turn. The upshot of all this is that people are less likely to engage with tweets over a certain age so the scope for the interaction to be sustained across a group of interested parties is limited by the scope that exists for related tweets to be encountered.

Certain mechanisms such as mentioning and favoriting provide a way for contributions by others to be highlighted within a user's own interface, and the use of hashtags provides a way in which users may at least return to a specific topic and explore what has lately been said. Furthermore, topic reference markers, as well as other strategies such as retweeting, can all be seen as ways in which people may attempt to enter or even reanimate the interaction at a later stage. Nonetheless, the fact remains that the structure of Twitter promotes engagement with tweets that are more recent within the overall stream and works against engagement with an unfolding interaction that has lapsed such that no contributions are visible from anyone you follow within the past few hours.

The core points to take note of here are that: i) the length of Twitter interactions is not pre-specified; ii) a range of techniques exists whereby Twitter interactions can be engaged in asynchronously and extended (potentially indefinitely) over time.

*5.1.9. The Content of Exchange.* A fundamental observation present in the *Simplest Systematics* is that 'what parties say is not specified in advance'. This can be seen to equally hold in Twitter-based exchange. A further, notable feature of Twitter interactions, is that they do not typically exhibit any systematic production of greeting, parting, opening or closing phenomena It needs to be remembered here that people producing turns in Twitter understand and therefore orient to their tweets as productions that recipients will encounter as and when they themselves choose to inspect the timeline. This being the case there is no need for work to be undertaken to establish a specific and contiguous space for the turn within ongoing interaction. Instead, this is provided for by a presumption of unspecified asynchronicity.

Additionally, because Twitter has an in principle right for self-selection when producing turns there is little work for greetings and partings to do, unless direct address to a specific follower or groups of followers is undertaken (as in Figs. 6a & b).

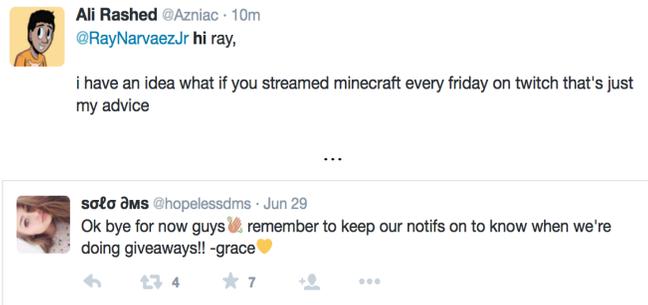

Figs. 6a & b. Openings and closing directed to specific recipients.

Furthermore, as we noted above many tweets are not produced with a built-in assumption that they will be productive of interaction anyway.

The key organizational feature to be noted here is that, along with many non-scripted turn-taking systems, Twitter exchange has the characteristic that the *content* of tweets is not specified in advance, even if certain elements (such as retweets and mentions) may be governed by certain conventions when they arise as *part* of the content.

*5.1.10. Distribution of Turns.* An interesting thing with Twitter is that one can inspect Twitter interactions and see systematically that the number of turns different contributors make can vary but without there being any evident externally imposed pattern. Instead you get people all jumping in separately to respond on occasion and just one or two on other occasions, even across the same groups of followers. Sometimes direct mentions and topical coherence will shape an interaction in such a way that it is, to all intents and purposes, purely dyadic. A critical thing to note here is that, even though there is a sense in which systematic features that have arisen in conversation to handle the problem of gaps and overlaps and ensure that there is only one speaker speaking at a time, are drawn upon as a matter of course in Twitter as well, the fact remains that the technology itself *overrides* any possibility of overlap and continually *imposes* gaps. Thus there is an orientation to managing an unfolding series of turns in microblogging that works as much as anything because of its to-hand nature (in that everyone already knew how to do conversation before any microblogging came on the scene) and its ready intelligibility to just anyone else. However, just because there are superficial similarities in the adoption of techniques, it should not be assumed that this therefore gives licence to treat Twitter-based exchanges as just another species of conversation and to therefore subject it to the conventions of traditional conversation analysis. Indeed, we have already seen that apparently similar techniques may be used in Twitter to achieve quite distinct ends.

The core observation to be made here is that there is no pre-defined distribution of tweets and turns at tweeting on Twitter. However, an important *outcome* of this feature is that, because of the technological constitution of Twitter and how this renders contributions to the timeline available to other viable recipients, Twitter interactions have a fragmented and non-contiguously paired appearance to recipients that makes them quite unlike face-to-face conversation.

*5.1.11. The Number of Potential Turn-Takers and Tweeter Selection (ii).* Something that Sacks et al. point to as systematically significant in the *Simplest Systematics* is the way in which the number of people involved in naturally-occurring face-to-face conversation is not pre-specified and can vary across the duration of a single conversation. This can be set against certain other kinds of spoken turn-taking systems where the number of parties who may participate can be very tightly specified, for instance during legal proceedings, during scripted activities such as rituals or theatre pieces, even during traditional telephone conversations where interactions are strictly dyadic, and similar kinds of two-way radio-based interactions.

It is equally the case in Twitter-based exchanges that the number of parties can vary systematically and without pre-specification. The only potential exception here is direct messaging where the interaction is dyadic in much the same way as it is for telephone calls etc. Indeed, just as telephone-based interaction can be observed to have some specific organizational features that set it apart from face-to-face conversational turn-taking, so one might say that direct messaging in Twitter (or in other microblogging domains) is a specialization of more general microblogging systems. Thus one can see in Twitter exchanges that remain more or less dyadic even though they are putatively public (an example of a dyadic Twitter interaction can be seen below in Figure 8), and on other occasions, exchanges that have exceptionally large numbers of contributors. A single recent interaction about the color of a dress on Twitter, for instance, attracted more than 650,000 responses. In fact, whilst the operating constraints of audibility and witnessable speaker selection serve to place a definite upper limit on the number of participants in ordinary face-to-face conversation, in the case of public microblogging domains the potential number of participants is constrained only by the extent of the follower network of the contributors to the interaction, making the possible upper limit potentially millions if the interaction involves a number of people with very large numbers of followers. Furthermore, it is clearly the case that on Twitter you can catch up on an exchange at any point (the tweets are there and you can read them at any point after the interaction has started). In spoken conversation, if you miss the beginning the scope to participate is extremely limited. The extent of visibility of Twitter interactions and the scope to contribute obviously also has implications for the possible transmission of rumors in public microblogging domains in that it can be systematically appropriate for large numbers of people to take a turn.

The critical point to be made here is that the number of potential parties involved in Twitter interactions is both varied and constrained only by the network of followers who might encounter a tweet (either in its original form or as a retweet)[15]. Thus there is an implicit understanding when one tweets that all followers may potentially respond. So, unless you are direct messaging, you cannot initiate a Twitter interaction with any assumption of it being dyadic. As we point out elsewhere in this paper, there *are* participant selection techniques available such as replying and mentioning but these do not close down the scope for other contributions in at all the same way as they would in face-to-face conversation. Instead the dominant mechanism for participant selection in Twitter is *self*-selection. Furthermore, it should be stressed here that the strong counterpart to the right to self-select in Twitter interactions is that one may also choose *not* to self-select. That is, one may encounter tweets and interactions yet not produce any kind of response. People do, of course, 'sit in' on face-to-face conversations in which they have no interest and may, for the larger part, 'zone out' of those conversations and play no part in them. However, in that they are present and can be understood to have heard what was said, they may nonetheless be accountably called upon to contribute, something most people have experienced from time to time with much accompanying embarrassment. No such accountable presumption is in play on Twitter and, across most microblogging sites, *non-selection is by far the most common phenomenon.*

---

[15] It is, of course, the case that Twitter also incorporates a search mechanism. For this reason, and in view of the fact that all tweets are essentially public outside of direct messages, any Twitter user can see any tweet in principle if they happen to use a search term that will reveal it. This is particularly the case during instances of significant breaking news, when large numbers of people may search on specific hashtag terms to try and find out more information. Generally, searching on hashtags provides an alternative mechanism for display to that conventionally bound up with friend-follower networks, though this of course in no way changes the core observation here that there are few constraints in place regarding the number of potential recipients of a tweet. We thank one of the reviewers of this paper for this useful additional observation.

*5.1.12. Continuity and Discontinuity.* Twitter is, by nature, synchronous or asynchronous in terms of response, with it usually being the case that a large number of unrelated tweets appear moment by moment within the stream, whilst tweets addressed to the same topic may potentially be widely spaced apart. A temporal consequence of this characteristic of Twitter is that the time spans over which respondents may address themselves to a topic without loss of coherence are much greater than they would be in face-to-face conversation.

The temporal organisation of Twitter means that there are certain distinct but equally systematic ways of marking out topic relationships that people will use in various sophisticated ways in order to manage coherence across more extended interactional threads. Re-tweeting is one obvious way in which this is accomplished.

Another more specific technique can be the use of the mention convention, which has the dual effect of both indicating the presence of a topic relation to all witnessing parties and of ensuring that the person specifically addressed sees one's tweet (Fig. 7).

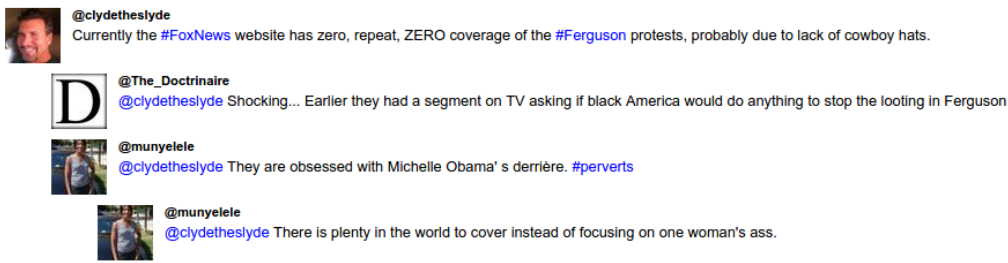

**Fig**. 7. **Use of the mention convention in Twitter.**

The organisation of conversation around topics, topical coherence, and shifts of topic is a central focus of the Conversation Analytic literature. Clearly responding to other people's tweets and using mentions as in the above, commenting upon embedded tweets being retweeted, and simple retweets all exhibit certain features of topical coherence, and Twitter itself also reflects this understanding in its grouping together of connected tweets in this way as 'conversations'. Grosser degrees of topical relation may also sometimes be encapsulated within the use of hashtags. At the same time, there are other indicators of 'on topic' / 'off topic' that can be seen to have a clear continuity with methods used in face-to-face conversation. For instance, note the use of 'btw' in the following and how the respondent handles both continuation of topic and the transition that has been proposed (Fig. 8).

Fig. 8. Managing change of topic in Twitter.

Principal matters to be attended to here are that tweets can be simultaneous, synchronous or asynchronous in terms of their composition, but are always either synchronous or asynchronous in their presentation on the timeline, with asychronicity being the norm. Key outcomes of this are that topical coherence in Twitter has less dependency upon adjacency than it does in many speech-based turn-taking systems such as face-to-face conversation. Nonetheless, Twitter also retains topic reference markers one *can* find in co-present interaction (e.g. 'as I was saying', 'btw'), as well as a number of more specialised techniques (e.g. the use of mentioning and hashtags).

*5.1.13. The Allocation of Turns and Tweeter Selection (iii).* Despite its largely asynchronous character and the potential interleaving of a number of distinct sequences of tweets on Twitter there are a number of turn allocation techniques can be observed. In the first instance, tweets are composed and arrive as distinct units within global Twitter feeds. With regard to any one particular topic being posited within these there is what might be considered to be a 'first speaker', from here on to be termed an 'originator', there are subsequent parties who may be implicated as respondents within the original tweet, and there are parties who select themselves as respondents to a tweet in some way. As indicated above, an important difference from face-to-face conversation here is the matter of 'co-placement', where responses to a specific tweet may not be sequentially directly adjacent to that tweet within a feed (because, in principal, all comers may respond to all tweets, so next up in a feed may be an entirely unrelated response to a different topic). For the allocation of turns it is also important to note that 'rights of response' in Twitter-based exchanges work in quite a distinct way. In principal any recipient of a tweet may respond to it or retweet it. This is not at all the case in face-to-face conversation, where just who gets to speak is a very tightly managed affair.

*5.1.14. Turn Constructional Units.* Something Sacks et al. make much of in the *Simplest Systematics* is the projectable character of just where a turn in talk might end. Twitter is evidently distinct here in a number of regards. First of all, because turns are usually already asynchronous, there is no need to provide for minimizing the gap between turns. At the same time. as we have already observed, there is an absolute length of 140 characters that constrains how long any 'constructional unit' might be. Nonetheless, if you can fit more than one sentence in your 140 characters that's absolutely fine and not a matter that is called to account by others. The example in Fig. 9 is just the originating tweet we saw in Fig. 1. Reference to Fig. 1 will show a number of other multi-sentence turns in that particular thread.

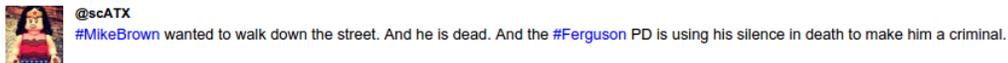

Fig. 9. A multi-sentence turn.

This scope for multi-sentence turns has structural consequences. For a start there is no need to project the ending. Recipients can see the whole turn as it has been crafted over any possible amount of time between previous postings. Additionally, recipients also understand that, without other indicators (such as we saw above regarding connected tweet markers), a tweet counts as the whole turn at talk and can be treated accordingly. Originators are thus accountable for its production and it is accountably appropriate to treat it as a complete turn even if it was posted prematurely by mistake. The fundamental point here is that turn-constructional units in Twitter are individual tweets, even if linking strategies are adopted. Furthermore, there is no scope for retrenchment, modification and repair within the tweet itself, something that happens routinely in face-to-face conversation as potential responses are foreseen and headed off before they become manifest. This matter of repair is an important aspect of any effective system for the organization of interaction.

*5.1.15. Provision for Repair.* In view of its importance, the organizational arrangements of tweet-exchange should clearly exhibit procedures for bringing about repair. It is of course the case that some of the technical aspects of the production of tweets are managed by the technology and not susceptible to user intervention or repair. In fact, a number of accounts and repair mechanisms are specifically directed at this characteristic of Twitter. For instance accounts may be directed to machine activity over which one has little control (Fig. 10).

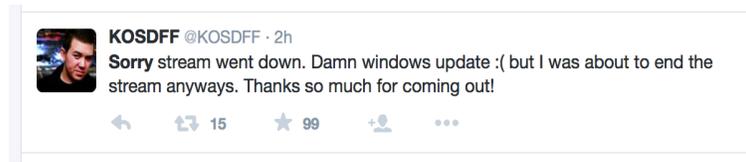

Fig. 10. Apologising for errors beyond one's control.

Other accounts, however, are framed in terms of user errors of one kind or another. Many accounts here can be found to relate to simple absence, as in the following. Note here how this provokes a further suggestion from the user that the absence of response is deliberate (Fig. 11):

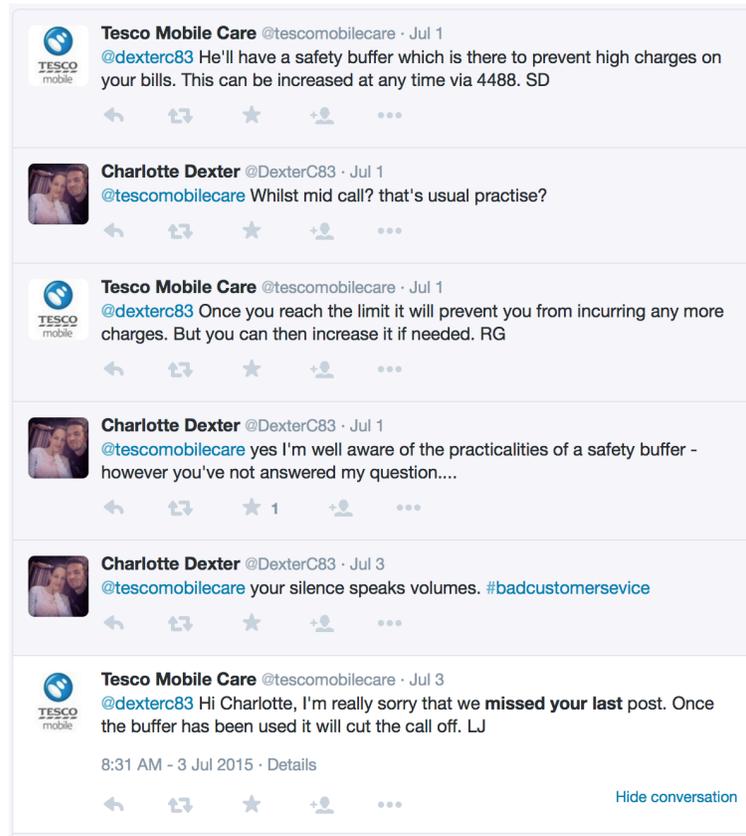

Fig. 11: Missing a post and being understood to be ignoring a complaint.

Another commonplace account is typing errors (Fig. 12):

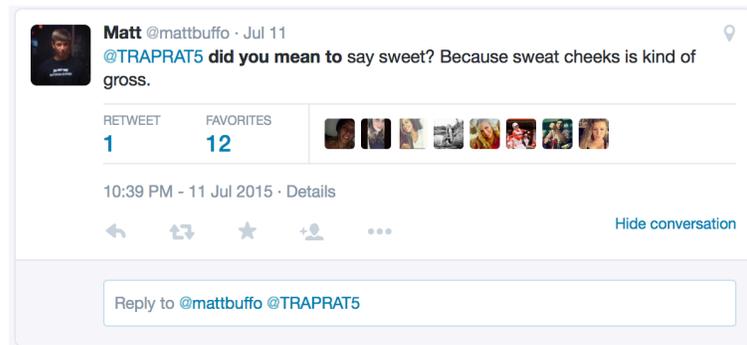

Fig. 12. Being pulled up on keying errors.

Yet another is lack of competence, such as the misuse of the mention facility (Fig. 13):

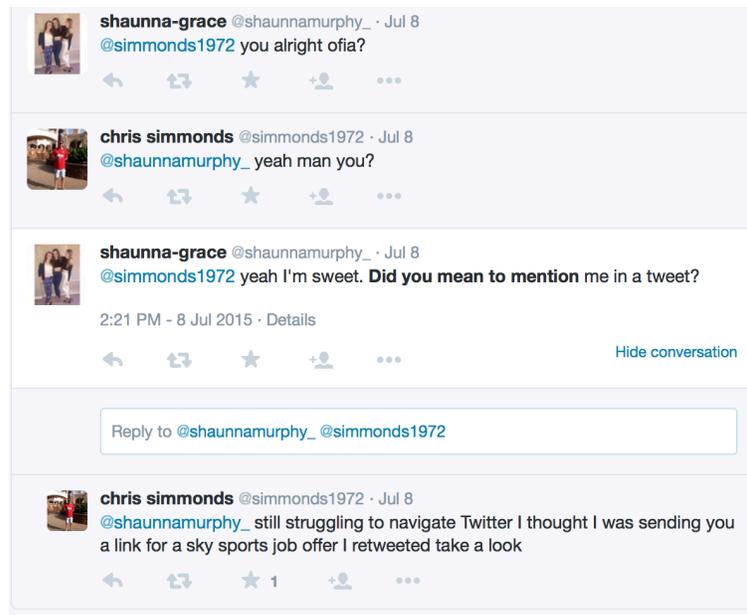

Fig. 13. Using mention by mistake.

People may also be pulled up on erroneous retweeting (Fig. 14).

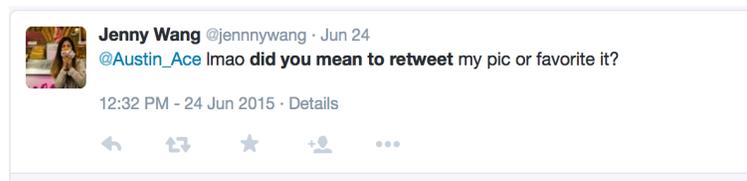

Fig. 14. Retweeting by mistake.

An important part of how tweeters are held accountable and their actions subject to repair relates to the *moral probity* of their actions. In this case Twitter is quite clear about the rights of users to effect a sort of repair by removing or deleting a tweet:

"Did you tweet something and then change your mind? Don't worry!
It's easy to delete one of your Tweets. Please note that you can only delete Tweets that you have made, you cannot delete other users'
Tweets from your timeline." [Twitter Help Center: 'Deleting a Tweet']

However, Twitter also makes the following observation:

**"Note:** Deleted Tweets sometimes hang out in Twitter search, they
will clear with time." [ibid.]

Additionally, there is nothing to stop users saving and then re-posting your
deleted tweet (although, on occasion, Twitter may treat the re-posting as a violation).
This can result in attempts at repair being unsuccessful (Fig. 15).

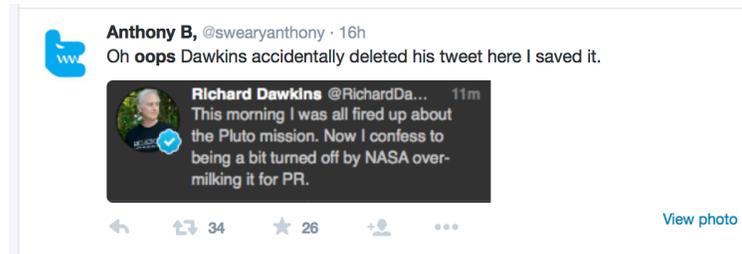

Fig. 15. Broadcasting a deleted tweet.

It should also be noted that Twitter does operate mechanisms for flagging offensive
tweets which may then be removed by Twitter itself [see Twitter Help Centre, The
Twitter Rules[16]]. Of course, there are occasions when users attempt to repair directly
misunderstandings that may have arisen regarding their posts (Fig. 16).

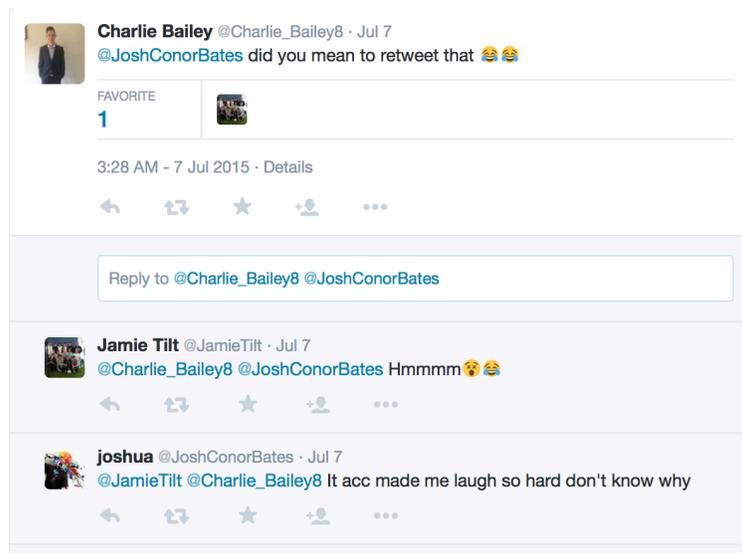

Fig. 16. Trying to account for an 'inappropriate' retweet.

And on some occasions tweets are called to account by others but nonetheless left to
stand. In the following an accusation of plagiarism provokes no response (Fig. 17):

16      https://support.twitter.com/articles/18311

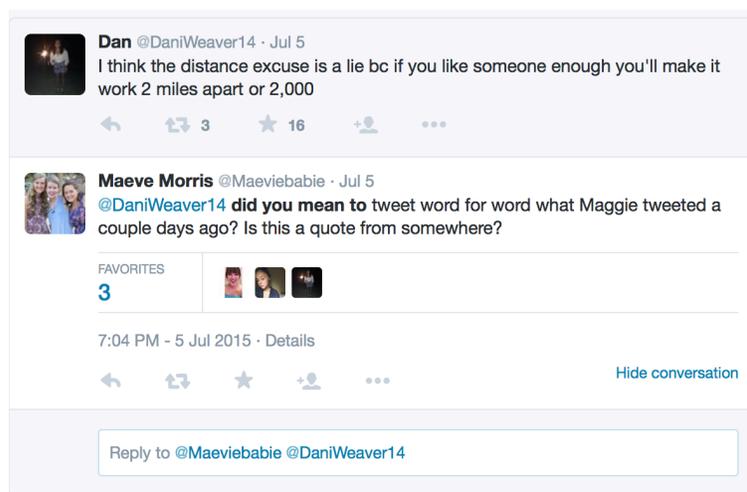

Fig. 17. Leaving a call to account unanswered.

An inspection of Twitter shows that a great many calls to account go unanswered in this way. The scope to override calls to account and expectations of repair in Twitter massively differs in this turn-taking environment from instances of face-to-face conversation. In face-to-face interaction calls to account are implicative for the production of the required account in the very next and adjacent turn. There are a number of foundational things to note about all of this: 1) There remains an understanding of how appropriate turns should be produced in Twitter; 2) There is a manifest orientation on the part of users to call other users to account when this commonsense order is breached in some way; *however,* 3) The strong orientation to therefore producing what in face-to-face conversation would be an expected second part in terms of an account or a repair of some kind is not similarly present. In consideration of this we should remind ourselves of the ways in which the turn-taking mechanism as described by Sacks et al. is powerfully constructed around the need to manage *co-present talk* so as to avoid unnecessary *gaps* or *overlaps* and an effective *distribution of turns*. We have already noted that Twitter's asynchronous character and absolute limit on the size of a possible turn, together with general rights of response, renders these kinds of management concerns redundant. As Twitter-based exchanges can occur between people who are otherwise strangers and who may never have another reason to interact, the scope for future face-to-face 'calling to account' is also minimal. Thus ignoring one's accountability for the turns one produces is much more likely to occur. This, too, has implications for how dispreferred actions such as the spreading of rumors may be more easily enacted via Twitter.

*5.1.16. Gap Management and Handling Asynchronous Interchange.* For Sacks et al. the fundamental concern was to arrive at a specific turn-taking model, or 'simplest systematics' for conversation. As the above review of how Twitter may operate as a turn-taking systems demonstrates, there are crucial differences between the model Sacks et al. came up with and what happens in Twitter. A key component of how these differences arise relates to how much of the turn-taking system in face-to-face conversation is pitched towards *minimizing gap and overlap*. Face-to-face conversation unfolds in co-present and linearly conjoint interaction such that gaps and overlaps are disruptive to the effective realisation of conversational talk. Tweets, by contrast, are textual productions that are, by virtue of the technical apparatus that enables them to be produced, both contiguous and without overlap. Thus *this is not a problem to which the construction of tweets needs to be addressed.* What one does encounter in Twitter, in particular in the context of what might be assembled by Twitter itself as a 'conversation', are phenomena such as 'the complete absence of a turn', that is to say a turn by a certain party may be projectible but not forthcoming. One may also encounter 'the conjoint production of largely unrelated turns'. That is to say, two (or more) followers may set out to respond to a prior tweet simultaneously in distinct ways, with the construction of Twitter resulting in their posts being assembled in the timeline consecutively even though they have no relation at all to each other but only to the prior turn[17]. In Fig. 18 below we can see how both @jawadmnazir and @flyhellas respond in very quick succession after the posting from @flightradar24 but in quite distinct ways. Furthermore, @flightradar24 only chooses to respond to the tweet from @flyhellas even though the questions posed by @jawadmnazir might be seen to be equally implicative.

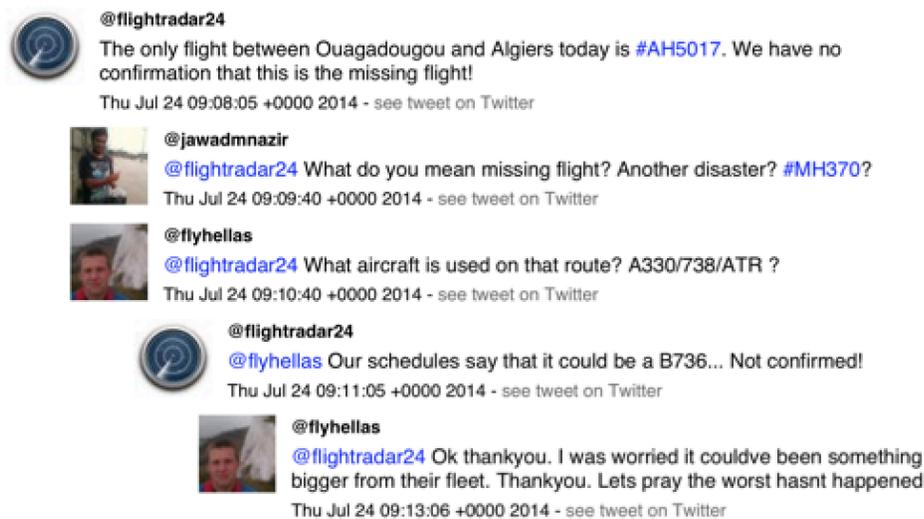

**Fig. 18. Consecutive disjuncture but mutual address to a preceding Tweet, together with disregarded implicativeness in Twitter.**

It thus falls to the recipients of the exchange to disambiguate the relationship of the various turns to one another without the availability of their sequential production standing as a resource for such disambiguation (as it would in conversation[18]). The only resource available in this regard is the gross fact that certain turns can be seen to precede others and that a turn will, by necessity, be addressed to some other turn that precedes it rather than to one that comes after it in the timeline.

---

[17]    Whilst it is organizationally different from this in a number of respects, Sacks et al. [1974: 712] do observe that the model they are proposing is foundationally geared to turn-taking in dyadic conversation with just two parties and that the addition of other parties can ramify. One of the ramifications they point to is that, when there are four or more parties, the talk can split up into more than one concurrent conversation with divergent talk happening at the same moment in time.

[18]    Note that, whilst it can also occur in face-to-face conversation that a number of parties may jump in at once with questions for somebody and a response be provided to only one of those questions in the first instance, this feature of interaction is well understood by all parties as a limitation imposed by the sequential production of speech. The key difference is that in spoken interaction the person being questioned remains accountable for the production of an answer to all of the questions posed and the questioning parties therefore have an unquestioned right to re-articulate their question until they do receive a response. This is not something we see happening much in Twitter.

The importance of this distinction between Twitter-based exchange and co-present conversation needs to be stressed. When Sacks et al. [1974:715] delve into part of the issue of why talking at once might be a problem and how the turn-taking system provides an economical method for handling that problem, they discuss in particular how the model, by providing for the analyzability of a turn of talk over the course of its production, might be impaired if turns were allowed to overlap, making projection of completion difficult to accomplish. Twitter has effectively obviated the need for its turn-taking system to handle this kind of problem by making it technically impossible for there to be overlapping turns.

However, it should be noted here that Twitter use has moved beyond how it was originally conceived by its creators as an apparatus for bringing about information exchange. Instead it is increasingly being oriented to as a device for specific user-to-user interaction over extended turns. This is recognised in the way Twitter itself now clusters related posts as 'conversations'[19]. This being the case the preceding observations can be seen to present a unique challenge to Twitter as it is currently constructed and bring into question the extent to which exchanges on Twitter can really be seen to operate as a conversations.

Another aspect of the *Simplest Systematics* is also crucially bound to the production of conversation in co-present interaction. This is the ongoing analysability of an utterance in the course of its production for its projectible point of completion[20] and for the kind of work that is being done. This enables a next turn-taker to be identified, for them to know when it is appropriate to take their turn, and for them to know what kind of a thing their own turn might need to accomplish[21]. Clearly, recipients of tweets are able to engage in a post-analysis of the whole turn at their leisure, even to the point of re-examining it multiple times, before they complete their reasoning about such matters, and without the pressure of needing to step straight in when up-and-coming completion of an utterance is recognised. This lack of in situ pressure to analyse and respond also renders tweeting distinct from certain other kinds of text exchange such as live chatting[22].

[19]      It should be noted that this feature of Twitter is still very limited in its application and we have uncovered numerous examples of clearly associated and 'interactionally-bound' tweets in our own research that were never actually represented as conversations within people's timelines.

[20]      Thus Sacks et al. [1974: 709] also note how variable turn-length is itself partly constituted by the nature of sentential constructions, which may themselves be extended through the inclusion of sub-clauses etc., and, in addition, comment that: *"Sentential constructions are capable of being analysed in the course of their production by a party/hearer able to use such analyses to project their possible directions and completion loci."*

[21]      Here Sacks et al. [1974: 710] point to the fact that, whilst 'what parties say is not specified in advance', certain kinds of turns do pre-figure what may thus be *done* with a subsequent turn, even if its exact content is not pre-specified. They additionally note that this feature can have an impact upon speaker selection in that certain types of turns pre-figure who the next speaker might be and what it is incumbent upon them to do.

[22]      Nonetheless, some Twitter exchanges do resemble live chatting with tweets passing to and fro in rapid succession. Although these clearly retain the structural features we are discussing there is obviously a need to respond to each other's tweets quite quickly or risk being understood to have 'left the conversation'. Thus it can be seen that there are actually a variety of different kinds of tweet exchanges on Twitter with different kinds of organizational expectations attached to them. These kinds of distinctions are one of the many things that an unfolding program of microblog analysis will need to encompass.

An overall summary of all of the above findings is presented below in Table 1, which provides an overview of the ways in which Twitter is distinctive as a turn-taking system.

| Turn-Taking Organizational Component | Specificity in Twitter |
|---|---|
| *Turn Change* | Turn-taking systems, by definition, require that turns be distributed amongst participants. The specific dynamic of this in Twitter is organized around follower/followed relationships. |
| *Turn Overlap & Turn Separation* | There is no *visible* turn overlap in Twitter. Within the interface all turns appear to be independent and consecutive. Distributed, text-based systems would all appear to share this characteristic. |
| *Gap Management* | Most turn-taking systems are organized to preserve turn continuity and minimize gaps between turns. This does not happen in Twitter because its technical structure prevents visible overlaps or gaps, whilst undermining continuity. In Twitter (and other microblogs) gaps between turns within a specific interactional sequence are the norm, but gaps in the visible presentation of turns are absent. |
| *Temporal Disjuncture* | Gaps of indeterminate length can happen between turns in Twitter. There is no indication of ongoing turn construction. Turns can be alternately realized with other activities. Turn delay does not manifest as a gap in the timeline, but rather as a delay in update. Turn delay is not oriented to as problematic. Co-present turn-taking systems are not delay-tolerant, but many online systems are. This does not apply to all text-based interactions equally. Most exhibit some tolerance for delay, but in online chat, SMS, etc., lengthy delay is accountable. |
| *Turn Order* | The order of turns in Twitter is not fixed. All followers have equal rights of response if a turn is not directed. Not all turns are implicative of a response. Turn order is dependent on the turn-taking system. Twitter and face-to-face conversation share flexible turn order, but operate at different scales, and non-implicative turns are rare in conversation. Non-implicative turns are also uncommon in online interaction such as online for forae and Q & A sites, and some turn-taking systems such as games and rituals have highly specified turn order. Non-implicative turns are, however, common across a range of microblogging platforms, making single turns the most common phenomenon. |
| *Turn Construction* | Twitter turns are tightly constrained at 140 characters, with a single tweet counting as a single turn. In synchronous and sequentially-bound turn-taking systems turn endings are projectible through their construction. This does not happen in Twitter (or other microblogs to any great extent). In spoken interaction turns are typically one-phrase long. Longer turns provide for seeing that a longer turn will happen. In Twitter and many other text-based forms, multi-phrase units are common. |
| *Turn Length* | Turn length in Twitter is tightly constrained at a 140-character maximum, though there are methods whereby the turn can be extended across multiple tweets. For many turn-taking systems, including other kinds of microblog, the length of the turn is not fixed and subject to significant variation. |
| *Turn Allocation* | Both implicated next turn-taker and self-selection for a turn happen in Twitter. This is similar to a number of turn-taking systems, including conversation. However, selecting the next turn-taker does not provide for that being the next visible turn and self-selection is far more common in microblogs. There is also no pre-defined distribution of turns in Twitter, such as one might find in rituals, games, and staged events. Any or all parties have a right to take a turn. Even face-to-face conversation is not as flexible on this score. |
| ***Participation Rights*** | Participation in Twitter is open in principle to any tweet recipient and therefore a direct function of who is being followed. This is the case with other microblog systems as well. In many co-present systems, such as conversation, despite the possibility of self-selection, not all parties will have equal rights to participate. This is even more the case with more formalized systems, such as games and rituals. A further distinction that especially holds for Twitter is that whole sequences of interaction may be inspected after the fact and then taken to be implicative for a new turn. This is only feasible in open text-based interaction, and, even then, in many cases there is a sense of an ongoing sequence that cannot be re-joined once it has expired. |
| ***Length of Exchange*** | For many turn-taking systems (rituals, games, enactments, etc.) the duration of the exchange is relatively fixed. Some systems, such as Twitter, other microblogs, online chat, and face-to-face conversation have no such constraint. Online interactions. especially microblog-based ones have the character of being open to continuation over extended periods as participants change location. This is not the case with conversation. Specific techniques have evolved in Twitter, such as hashtags, retweets and mentions, that can serve to support more extended courses of interaction. |
| ***Content of Exchange*** | There is no pre-specification of what the content of a turn should be in Twitter. This is the case across numerous turn-taking systems (excepting things like rituals, games, staged events, etc.). An added feature is that there are no specific opening or closing phenomena in Twitter, which renders it distinct from may other systems including online chat. This is, however, something that can be found in other microblogging systems. |
| ***Number of Interactants*** | The number of potential turn-takers in Twitter is not pre-specified. This is superficially similar to other systems such as conversation and differs from dyadic telephone calls, dyadic text such as messaging and formalized interaction such as rituals or staged events. However, the upper limit of participants in Twitter and other microblogs is limited only by the extent of follower/followed relations, whilst in situated interaction it is limited by the number of parties physically able to be co-present. |
| ***Continuity and Discontinuity*** | Continuity markers are present across numerous turn-taking systems. In Twitter, however, they are chiefly oriented to as a solution to the constrained length of tweets. In other systems they are more typically used to preserve topic coherence., though there are techniques for managing turns that are on/off topic in Twitter as well.<br><br>In Twitter discontinuity is the norm, with responses rarely falling adjacent to the prior turn within the tweetstream. This is a distinctive feature of microblogs, which provide systematically for the conjoint production of unrelated turns. In most turn-taking systems some kind of provision is made to preserve continuity across turns. |
| ***Provision for Repair*** | Turns in Twitter cannot be repaired as they are being made available to other participants. Repair has to be accomplished through a separate turn. In co-present systems a turn can be repaired over the source of its production.<br><br>Turns in Twitter can be called to account. However, lack of adjacency in response and reduced social obligation result in many calls to account going unanswered. Most turn-taking systems display stronger attention to the provision of accounts. Even other microblogs where followers are typically 'known' parties  show stronger recognition of accountability and obligations to repair. |

**Table 1. The distinctiveness of Twitter as a turn-taking system.**

A question that might be posed with regard to this is the extent to which a similar distinctiveness might be uncovered when examining the actual practices involved in using any specific system that is designed to support some kind of interaction. An important feature of the table is the extent to which it also makes visible the ways in which certain turn-taking elements are contiguous across a range of different turn-taking systems. However, it is also important to take care not to see this as an indication of where generalisability might be discovered. As we note, for instance, regarding the way in which different systems may still draw upon continuity markers to bind distinct turns together in the exchange, the use of these mechanisms is not necessarily oriented towards accomplishing the same ends in every case. Thus continuity markers in Twitter may also offer a solution to getting around the constrained length of a specific turn. The extent to which apparently shared mechanisms are being used to the same purpose is therefore a topic that should be investigated in further research.

*5.1.17. The Analytic Enterprise.* There is one further very important aspect of turn production that merits discussion here. Sacks et al. highlight it in their *Simplest Systematics* in the following way:

> "... while understandings of other turns' talk are displayed to co-participants, they are available as well to professional analysts, who are thereby afforded a proof criterion (and a search procedure) for the analysis of what a turn's talk is occupied with. Since it is the parties' understandings of prior turns' talk that is relevant to their construction of next turns, it is their understandings that are wanted for analysis. The display of those understandings in the talk of subsequent turns afford both a resource for the analysis of prior turns and a proof procedure for professional analyses of prior turns - resources intrinsic to the data themselves." [op cit: 725].

In other words, the local analysis of a prior turn that is made visible in a subsequent turn is itself a resource for our own analysis. It tells us what the participants to a course of interaction themselves understand to have been done at every step of the way. To return for a moment in that case to our founding interest in rumor propagation, we can observe that what counts as a rumor is what is manifestly taken to be a rumor and handled that way in the turn that follows what is seen to be the source of the rumor in the first place. Thus there is no point in looking to any one turn and seeing it as amounting to a rumor in any free-standing way. It is the turns that follow that will be seen to matter. So, in Fig. 19 the initial item of 'breaking news' would have remained just that – 'news' – if it had not received 112 retweets and had various comments made upon it that themselves made it a piece of putative information that was being shared with other people. And the clinching point was the moment when its status as something other than 'just news' – 'not according to what I've just heard on CTV' – made it open to being recognized by other parties as something unverified and therefore possibly 'just a rumor'.

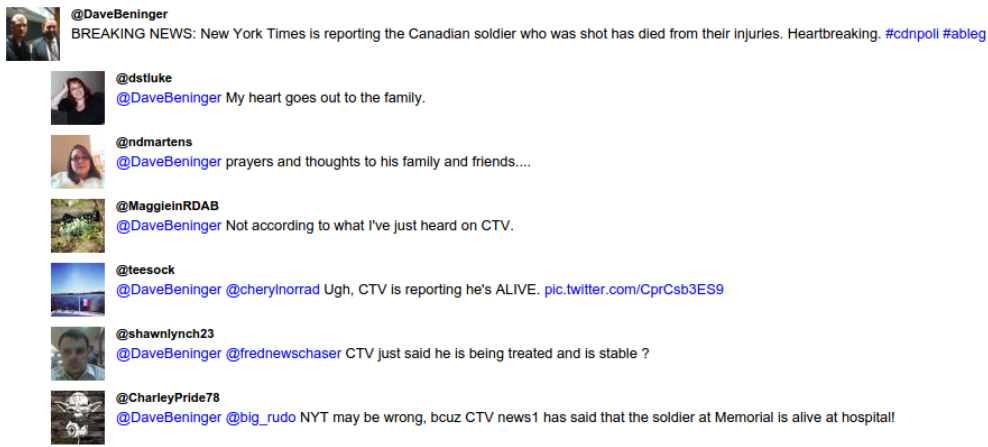

**Fig. 19. An unfolding rumor.**

## 6. APPLIED MICROBLOG ANALYSIS: EXAMINING RUMOR

The above insight that rumor is by necessity a collaborative production that unfolds across multiple turns in any exchange system has been central to our own development of a suitable annotation system for handling rumor production in Twitter. This point is foundational. One needs to examine the organizational characteristics of how specific phenomena unfold in social interaction to understand how they *work* as social phenomena. So it is only by examining how tweeting (or microblogging more generally) is socially organized, *in its own terms*, that one can begin to understand how any social phenomenon enacted through it is realized, including rumor. The insights from Sacks et al. we focused on above are not insights about conversation *per se*, even though Sacks et al. were concentrating upon conversational phenomena at the time. They amount to being insights about turn-taking systems in general, i.e.: turn-taking systems are fundamentally organized to be implicative; to allow interaction to unfold over time. This means that focusing upon any one isolated turn is very limited with regard to what it can tell you about how it may or may not be *socially* meaningful. This matters as much for Twitter as it does for face-to-face conversation. And it matters as much for rumor as for any other socially recognizable phenomenon in Twitter (joking, teasing, arguing, promising, requesting, telling, sharing, ranting, suggesting, and so on). As a concluding exercise we will therefore begin to outline some of the ways this approach has begun to be used specifically with regard to the production of rumors on Twitter, and how that in turn has begun to inform certain aspects of systems design.

The above discussion is heavily centered around sociological concerns and sociological analysis. In this final section, however, our aim is to illustrate that, despite this sociological emphasis, our interest is very much bound up with a desire to inform and assist systems design. In this regard it is wholly commensurate with concerns that are at the heart of Human-Computer Interaction as a domain. It is worth reminding the reader at this point that our core aim in examining Twitter as a socially constituted phenomenon has been to inform the development of an annotation scheme that we may then use to help solve the challenges of early detection and veracity evaluation of rumors in social media [Bontcheva et al., 2015; Derczynski et al., 2015; Lukasik et al., submitted; Zubiaga et al., 2016a; Zubiaga et al., 2016b]. The sociological exercise has been one of setting that endeavor upon a sound and rigorous methodological footing. To do that has entailed going back to first principals because we found existing schemes of analysis not yet quite fully attuned to the uniqueness of Twitter as a domain of social interaction.

Whilst there is much still to be done on both the analytic and design front, there are certain foundational elements we have already begun to pull out of the above observations and embed within our ongoing development of a suitable annotation scheme. These particularly relate to the following features of Twitter use:

—(1) That it is sequentially ordered

—(2) That Twitter- based exchanges involve topic management

—(3) That there are important accountability mechanisms in play

Additional features that are alluded to in the closing example above that are more specifically related to our interest in rumor production include matters such as:

—(4) Agreement and disagreement

—(5) The ways in which tweets are rendered trustworthy and believable through the production or otherwise of evidence

More specifically our annotation scheme for rumorous interactions reflects sequential ordering by providing for the annotation of triples of original and reply tweets, where the original tweet is visible as a topic reminder in nested replies. Annotation was originally undertaken using a crowdsourcing platform [Zubiaga et al, 2015b]. Annotators were asked to identify the ways in which source tweets either aligned with or refuted specific news items that might count as rumors (relating to (2) and (3) above). Subsequent tweets in the sequence were then annotated regarding the extent to which they either agreed or disagreed with the source tweet ((1), (2), (3) and (4) above). The tweets were then further annotated to indicate the degrees of certainty and use of supporting evidence within each post ((5) above). The crowdsourced annotations were then used to train algorithms that are able to automatically assign 'veracity scores' that make visible the system's confidence in the degree to which a given rumor might be true at any point within a particular thread's lifecycle. This has been mainly pitched towards the creation of a dashboard for journalists [Tolmie et al., 2017] with the goal being to provide journalists with in situ support that is based upon machine learning algorithms that have been initially trained on exactly the kinds of Twitter-based interactions journalists routinely draw upon in their everyday work.

So, in Figure 20 below we can see the following annotations being applied: The 'support' element of the initial source tweet is annotated as 'supporting' because the tweet does not seek to refute the claim presented within it. It is also annotated to show whether it provides any supporting evidence for the claim within the body of the tweet. Thus for 'evidentiality' the annotation notes that a URL is given. Assessment of the propositional orientation to the content of the tweet leads to the further annotation that the 'certainty' of the tweet is 'certain'. Subsequent tweets in the interchange (indicated by the indented nesting) are then annotated in the following kinds of ways: The orientation towards the initial tweet is captured by 'responsetype vs source'. These can be seen to be variously 'comment' or 'agreed', indicating that the respondents were either simply commenting on the content or actively agreeing with it. A further series of nested tweets capture the responses to the previous tweet, rather than the source, reflecting the concern we had with analyzing three-turn structures to understand how initial turns were being handled in actual interaction. These tweets are annotated in terms of 'responsetype vs previous', and it can be seen that these cover 'agreed', 'comment', or 'disagreed', to capture the degrees of agreement or disagreement of the response. These are additionally annotated with regard to the orientation they also display towards the original source tweet, this amounting to being either 'agreed' or 'comment'. Further annotations capture aspects such as 'evidentiality' (e.g 'witnessed' because the respondents have seen the pictures that are the cause of the debate), and levels of certainty in the propositional content of the tweet (e.g. 'certain').

For a more detailed description of the annotation scheme the reader is referred to [Zubiaga et al., 2015a; 2015b; 2016a]. What we would like to underscore here through the examination of a couple of examples of Twitter-based exchanges is the scope for the kind of analysis we have been articulating here to inform rich annotation of microblog materials towards a variety of ends where there is a need to understand people's situated practices, reasoning and *methods* for using microblogs.

**Fig. 20**. Example of a rumor where the truth status of the original post is not brought into question.

The interaction presented in Fig. 20 refers to a post releasing a set of photos taken from an eyewitness video of the Charlie Hebdo attacks in Paris. Here we see how the validity of the original post is not brought into question and the remaining posts take the form of commentary upon it.

At the most basic level we can see how this example is a series of turns with not just one tweeter doing all the work. In section 5.1.1 we note how this is the most basic requirement for a turn-taking system to function. So the post has implicated a response, which sets it aside from the many Twitter posts that are oriented to as free-standing. At the same time, it is only in the latter part that we can see any tweeters taking more than one turn. A thing we cannot see here because we do not have access to each individual party's timeline is the extent to which the turns in the interaction are appearing in a disjunct fashion, with other unrelated tweets appearing in-between, but the likelihood is that for most of these parties this is how the interaction would have appeared. We note in 5.1.4 that for most Twitter users posts even to the same conversation will appear in a disjoint fashion in their tweestreams unless they specifically choose to display the posts as a conversation. The exact spread here would have depended upon the number of other parties they were following and the degree to which those other parties were also tweeting at the time. There are an number of other routine features we can see here. Each tweet is complete as a turn without any overlap. We note in 5.1.3 that this is a given feature of Twitter because of the way it has been designed at a technical level. Similarly, as we note in 5.1.7, each turn is accomplished in under 140 characters. We also see no extension strategies being adopted here, such as the use of continuation dots or the numbering of posts. Several turns here also demonstrate the use of multi-sentence units, which. as we note in 5.1.14, is one of the ways in which turn-taking in Twitter escapes the confines of projectible completion points to be found in face-to-face conversation. Without the presence of timestamps[23] we cannot be certain how much temporal disjuncture there was between each of these tweets but, once again as we note in 5.1.4, any inspection of Twitter timelines would make clear the likelihood that this was the case. The lack of timestamps also makes it hard to assess the degree of temporal discontinuity within the interaction, but, as is discussed in 5.1.12, even short interactions such as this can unfold over hours as different respondents pick up the thread, making these kinds of interactions radically distinct from the temporally and sequentially-bound interactions you encounter in face-to-face conversation. What remains constant and binds together the interactional relation is the persistent mention '@independent', with a number of parties also including the picture link. Only one party here broadens out the topic relations within the thread by using a hashtag, but in some series of tweets one can see a hashtag in virtually every response (see Fig. 21).

[23] The absence of timestamps in this case was an artifact of the process of creating the visualizations, so close analysis of the exact temporal relationship between tweets would be feasible if that was critical for understanding the interactional order. For our own purposes, at this point in the development of a working system, this did not appear to have any bearing upon the annotation process. However, if the contrary later proves to be the case, timestamps will be a part of the visualization as well. Nonetheless, this does point to a larger issue regarding the conduct of the kind of analysis proposed in this paper. Actually collecting all aspects of every relevant post (or even every post) from a particular thread, including the original, is subject to the extent to which one has access to the platform, the kinds of requests the platform provider is willing to process, and the nature of the specific API you are obliged to use. Not all platform providers are equally open about this. In the case of Twitter a number of restrictions are in place regarding what kind of access they will give to stored tweets and certain kinds of access involve an expensive commercial transaction.

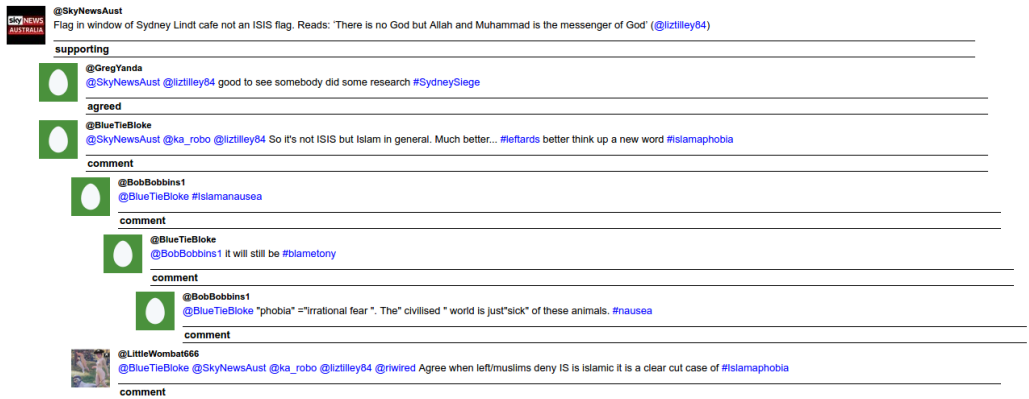

Fig. 21. Recurrent use of hashtags.

An additional routine feature we can note in the exchange in Figure 20 is the absence of greetings or partings or any formal closing down of the interactional thread. In section 5.1.9 we discuss how this absence of openings and closings is another distinct feature of microblog exchanges because posts are typically oriented to as turns that will be encountered by whomever, whenever. Thus first posts do not have any distinct structural work to accomplish, and it typically unknown whether your turn is going to be the last turn in the sequence. Beyond this, note the extent to which people assume without question their right to self-select in order to respond to prior tweets, with the exact order of turns and respondents being unspecified. We discuss in a number of places in section 5 the way that self-selection is commonplace in Twitter, rather than specific turns by specific parties being directly implicated. At the same time, there are a number of mentions without the next respondent turning out to be the party implicated by the mention. In 5.1.6, 5.1.11 and 5.1.13 above, we point out that, whilst the use of devices such as the mention in Twitter does provide for selecting the next party to take a turn, in practice the responses are rarely adjacent and the interactional sequence is often fragmented by other parties self-selecting before the selected party has had a turn. Formally implicated responses are only visible in the final 3 tweets of the thread. This underscores the extent to which the distribution of turns and number of participants can be variable, with an open series of responses, followed by a short, more dyadic form of interchange where direct mentions serve to implicate specific forms of response. At the same time we can see the absence of implicated response from @andremurphy and @independent. In 5.1.16 we discuss how an absence of response is commonplace in Twitter and is not apparently treated as being accountable in the way it would be in many other turn-taking systems. Another point of interest is how the tweeters involved are aligning to subtly distinct matters of topic, distinctions that are not wholly commensurate with the post relationships themselves. What we can see are three related but different concerns being addressed by the parties. Some are concerned to assign a moral ascription to the matter overall, e.g. 'terrible'. A second set of posts is concerned with the identity of the wounded man as a police officer. What particularly characterizes this set of tweets, however, is their organization around calling to account @independent for posting a picture of the dying police officer. In other words this sequence is focused upon a matter of moral probity and we can also see here how this calling to account actually goes unanswered. Once again, this is something that appears to be a common feature of accountability management in Twitter that is quite distinct from ordinary conversational callings to account, as noted in our discussion of repair in section 5.1.15. We have also pointed out that the apparently reduced concern with specific accountability is a feature that may well serve to shape Twitter-based exchanges as a particularly potent mechanism for rumor propagation.

**@DaveBeninger**
BREAKING NEWS: New York Times is reporting the Canadian soldier who was shot has died from their injuries. Heartbreaking. #cdnpoli #ableg
Annotation: certainty: certain, evidentiality: source-quoted, support: supporting.

**@dstluke**
@DaveBeninger My heart goes out to the family.
Annotation: responsetype-vs-source: comment.

**@ndmartens**
@DaveBeninger prayers and thoughts to his family and friends....
Annotation: responsetype-vs-source: comment.

**@MaggieinRDAB**
@DaveBeninger Not according to what I've just heard on CTV.
Annotation: certainty: somewhat-certain, evidentiality: no-evidence, responsetype-vs-source: disagreed.

**@teesock**
@DaveBeninger @cheryInorrad Ugh, CTV is reporting he's ALIVE. pic.twitter.com/CprCsb3E59
Annotation: certainty: certain, evidentiality: source-quoted, responsetype-vs-source: disagreed.

**@shawnlynch23**
@DaveBeninger @frednewschaser CTV just said he is being treated and is stable ?
Annotation: certainty: uncertain, evidentiality: source-quoted, responsetype-vs-source: disagreed.

**@CharleyPride78**
@DaveBeninger @big_rudo NYT may be wrong, bcuz CTV news1 has said that the soldier at Memorial is alive at hospital!
Annotation: certainty: somewhat-certain, evidentiality: source-quoted, responsetype-vs-source: disagreed.

**@loas_ia**
@DaveBeninger maybe you should get your info a little more locally
Annotation: responsetype-vs-source: comment.

**@reganoglivie**
@DaveBeninger @bebeasley Careful when reporting on life and death. Let's wait until we really know.. when the news is not so chaotic
Annotation: responsetype-vs-source: comment.

**@Donnajcherold**
@DaveBeninger @doctorfullerton no report he has died!
Annotation: certainty: uncertain, evidentiality: reasoning, responsetype-vs-source: disagreed.

**@doctorfullerton**
@Donnajcherold @DaveBeninger New report indicates he is alive but obviously gravely injured. I'm not sure where New York Times got that.
Annotation: certainty: somewhat-certain, evidentiality: no-evidence, responsetype-vs-source: disagreed, responsetype-vs-previous: agreed.

**@Donnajcherold**
@doctorfullerton @DaveBeninger me either. There are reports he is still alive. But not looking good! #OttawaShooting
Annotation: certainty: somewhat-certain, evidentiality: no-evidence, responsetype-vs-source: disagreed, responsetype-vs-previous: agreed.

**@NatriceR**
@Donnajcherold @doctorfullerton @DaveBeninger hi Donna I think we are talking 2 different incidents now the quebec one is what I referred 2
Annotation: responsetype-vs-source: comment, responsetype-vs-previous: comment.

**@Donnajcherold**
@NatriceR @doctorfullerton @DaveBeninger ??????
Annotation: certainty: uncertain, evidentiality: no-evidence, responsetype-vs-source: appeal-for-more-information, responsetype-vs-previous: appeal-for-more-information.

**@NatriceR**
@Donnajcherold @doctorfullerton @DaveBeninger Donna - the rcmp incident in Quebec versus today in ottawa
Annotation: responsetype-vs-source: comment, responsetype-vs-previous: comment.

**@doctorfullerton**
@NatriceR @Donnajcherold @DaveBeninger But both would indicate there is need to increase surveillance & ability to detain
Annotation: responsetype-vs-source: comment, responsetype-vs-previous: comment.

**@DaPro63**
@DaveBeninger @SheilaGunnReid Canadian news contradicts this
Annotation: certainty: uncertain, evidentiality: no-evidence, responsetype-vs-source: disagreed.

**@SheilaGunnReid**
@DaPro63 twitter.com/JohnW_MP/statu... @DaveBeninger
Annotation: responsetype-vs-source: comment, responsetype-vs-previous: comment.

**@DaPro63**
@SheilaGunnReid @DaveBeninger perhaps wishful thinking on my part but still hopeful it is not true. TV still not reporting any CF deaths
Annotation: responsetype-vs-source: comment, responsetype-vs-previous: comment.

**@TimothyEWilson**
@DaveBeninger @SheilaGunnReid Perhaps wait for Canadian source.
Annotation: certainty: uncertain, evidentiality: no-evidence, responsetype-vs-source: appeal-for-more-information.

**@SheilaGunnReid**
@TimothyEWilson twitter.com/JohnW_MP/statu... @DaveBeninger
Annotation: responsetype-vs-source: comment, responsetype-vs-previous: comment.

**Fig. 22. Example of a false rumor.**

The Twitter thread in Fig. 22 demonstrates many of the same kinds of routine organizational features that we saw in the preceding example: frequent tweeter change and turn allocation on the basis of self-selection; tweet separation and potential temporal disjuncture; the use of mentions to provide for topical coherence and to implicate further turns; and, once again notably, the absence of response from the originator to a calling to account, this time explicitly for possible rumor-mongering.

The sequence of tweets tackles the prospectively rumorous post from a variety of perspectives, displaying a number of ways in which accountability mechanisms may be visibly brought to bear upon unfolding content of this kind. Ultimately, it turns out that the foundation of the post as a 'false' rumor hinges upon a confusion of events. The initial post cites the New York Times as saying the Canadian soldier shot in the Ottawa shootings has died. Responses to this initially don't bring it into question and instead align with content in ways that are similar to the example in Figure 20. However, a post by @CharleyPride78 then enters the timeline, saying that a Canadian TV station is reporting that the soldier is alive. There are then numerous posts aligning with this post, some of which call the original tweeter to account for having posted false information. It is only towards the end that a post by @NatriceR suggests the possibility that there has been a confusion of events with the death of the soldier referring to an earlier event in Quebec instead.

The primary cohering feature in this second example is the mention throughout of the original tweeter, @DaveBeninger, which reflects how we saw the mention of @independent being used in the preceding example. Groups of tweets within the exchange, however, then cohere around a range of other mentions, not all of whom are even visible within the timeline, possibly because they are simply retweeting the posts of others, e.g.: @cherylnorrad; @frednewschaser; @big rudo; @bebeasley; and so on. We note in section 5.1.5 how retweeting can serve to potentially extend the cohort of possible participants indefinitely, a feature that is highly distinct to Twitter because of the open nature of its follower relationships. A matter of potential importance here is the kinds of considerations being brought to bear by people when they use mentions within interactional streams like this. The mention can be serving to both evidence response and to implicate further responses form others, making it a potentially powerful turn-taking mechanism. As is noted in our discussion of our annotation scheme, this is also a way in which interactants can display degrees of alignment and make manifest certain kinds of evidence. Another feature of interest within this example is the distinct way in which non-alignment and dispute of the original tweet gets marked out within the textual realization of the response, with certain markers of disagreement one might also find in face-to-face conversation, e.g. the 'ugh' in the 5th tweet in the thread. However, in ordinary conversation these kinds of markers provide for the seeability of an up-and-coming dispreferred response and give the originator an opportunity to engage in repair. In Twitter, as we have now seen across several different examples, accountability and repair mechanisms do not necessarily operate in the same kinds of ways and the use of disagreement markers cannot therefore be presumed to work in the same way either (see once again 5.1.15 above). Indeed, as we have already noted, at no point does @DaveBeninger actually respond.

## 7. CONCLUSION

In this paper we have set out to explicate how the principles underlying conversational analytic approaches are also fundamental to understanding the social organization of microblogging in platforms such as Twitter. It quickly became apparent, however, that to just apply the apparatus of conversation analysis was not wholly satisfactory; microblogging, for all of its conversational characteristics, is not conversation. Hence, our focus has been to unpack some of the important distinctions between conversation and microblogging. Building on this, we have proposed a programme of analysis that is better suited to microblogging and we have used the example of an annotation scheme we have developed for rumor to illustrate some of the ways this might then be applied. Of course, our argument is that this approach, including the annotation scheme, is applicable beyond studies of rumor to a wider range of phenomena inherent in social media-based interactions. In support of this, we would note that this emphasis on the social organization of microblogging is consistent with the emergence of machine learning social media analytics that focus not on individual postings but on the threads of which they are manifestly part (Ritter et al. [2010]; Schantl et al. [2013]; Bak et al. [2014]; Zarisheva and Scheffler [2015]; Webb et al. [2016]; Housley et al. [submitted]).

Clearly, there is much still to be done with regard to both the development of microblog analysis and attached design endeavors such as the annotation scheme relating to rumors. This paper should therefore be understood as an initial foray into the landscape that attempts to lay out some of the most important foundational considerations and organizational features, exploring the most effective means for what Garfinkel [1967] calls finding 'the animal in the foliage'. Our own work has centred largely on applied cases of microblogging where rumor production and veracity is of central concern, such as journalism, to date (Tolmie, et al., [2017]), though we are looking to other domains to extend our interest. Hopefully, however, the reader can begin to grasp the key point here that understanding how any of the technological components of Twitter or any other kind of microblogging can have any kind of *social* import or meaning turns upon understanding how they are socially organized productions. This social organization is to be found in the detail of just how their exchanges are brought about and subjected to ordinary everyday assumptions and reasoning. And this body of reasoning has to be understood in its own right, not simply as a subspecies of face-to-face conversation. More than this, we have already outlined above how sociological analysis of this kind is not simply an arcane pursuit but rather at the very heart of social computing and the use of studies of human interaction to inform systems design. The work we have begun on the discovery and annotation of rumors is an instantiation of exactly this concern with embedding systems design in a rigorous understanding of how the social world is accomplished. This, we argue, is especially important for digital technologies that are inherently social such that users are in a position to play a critical role in shaping them; that is they are 'co-produced' by the activities of their users. We would therefore hope that the insights provided in this paper and, most especially, the *approach* might be taken on by other researchers who are keen to unpack Twitter-based phenomena to inform design across a broad spectrum of interests. Furthermore, we would hope that the programmatic character of the approach we are advocating might serve to inspire researchers examining other co-produced technologies, including not only other forms of microblogging such as Facebook, Tumblr, or Reddit, but also other socially constituted online resources for interaction such as Skype, the vast array of worlds dedicated to online gaming, etc., not to mention the innumerable other such technologies that have yet to put in an appearance.

## ACKNOWLEDGMENTS

The research reported in this paper is supported by the EC FP7-ICT Collaborative Project PHEME (No. 611233).